\documentclass[12pt]{article}

\usepackage{amssymb,amsmath,amsfonts,eurosym,geometry,ulem,graphicx,caption,color,setspace,sectsty,comment,footmisc,caption,pdflscape,subfigure,array,hyperref,pgf,tikz, rotating, float}
\usetikzlibrary{positioning}
\usepackage[round]{natbib}
\usepackage[flushleft]{threeparttable}
\usepackage{adjustbox}
\usepackage{rotating}

\newcolumntype{L}[1]{>{\raggedright\let\newline\\arraybackslash\hspace{0pt}}m{#1}}
\newcolumntype{C}[1]{>{\centering\let\newline\\arraybackslash\hspace{0pt}}m{#1}}
\newcolumntype{R}[1]{>{\raggedleft\let\newline\\arraybackslash\hspace{0pt}}m{#1}}

\newcommand*\mycirc[1]{%
  \begin{tikzpicture}
    \node[draw,circle,inner sep=1pt] {#1};
  \end{tikzpicture}}

\begin{document}

\begin{titlepage}
\title{The Impacts of Mobility on Covid-19 Dynamics: Using Soft and Hard Data}
\author{Leonardo Martins\thanks{Department of Economics, Pontifical Catholic University of Rio de Janeiro - Brazil. Email: lcladalardo@gmail.com} \and Marcelo Medeiros\thanks{Department of Economics, Pontifical Catholic University of Rio de Janeiro - Brazil. Email: mcm@econ.puc-rio.br}}
\date{\today}

\maketitle

\begin{abstract}
\noindent This paper has the goal of evaluating how changes in mobility has affected the infection spread of Covid-19 throughout the 2020-2021 years. However, identifying a ``clean" causal relation is not an easy task due to a high number of non-observable (behavioral) effects. We suggest the usage of Google Trends and News-based indexes as controls for some of these behavioral effects and we find that a 1\% increase in residential mobility (i.e. a reduction in overall mobility) have significant impacts for reducing both Covid-19 cases (at least 3.02\% on a one-month horizon) and deaths (at least 2.43\% at the two-weeks horizon) over the 2020-2021 sample. We also evaluate the effects of mobility on Covid-19 spread on the restricted sample (only 2020) where vaccines were not available. The results of diminishing mobility over cases and deaths on the restricted sample are still observable (with similar magnitudes in terms of residential mobility) and cumulative higher, as the effects of restricting workplace mobility turns to be also significant: a 1\% decrease in workplace mobility diminishes cases around 1\% and deaths around 2\%.
\vspace{0in}
\vspace{0in}

\bigskip
\end{abstract}
\setcounter{page}{0}
\thispagestyle{empty}
\end{titlepage}
\pagebreak \newpage

\doublespacing

\section{Introduction} \label{sec:introduction}

The Covid-19 pandemic has created a new dynamic in terms of social behavior. Its impacts over society are widespread through all fields, from psychological effects on individuals (as in \cite{kontoangelos2020mental}) up to economic effects over countries,  as examined by International Monetary Fund (IMF) publications  (e.g. \cite{deb2020economic}). In this paper we aim at a particular effect of the virus-spread: the impacts of restrictions to mobility over its effects on Covid-19 cases and deaths. We use a panel dataset at the municipal level in Brazil to measure the short-term impacts of reduction in mobility on the dynamics of Covid-19.

Throughout the first year of the pandemic (2020), many countries adopted circulation restrictions with the objective of reducing the spread of the disease on the population. However, there is a large heterogeneity in terms of the restriction degrees over the countries: while New Zealand imposed a high-level centralized lockdown strategy, Brazil only imposed decentralized mobility restrictions. In this scenario, evaluating causal effects of mobility impacts on the infection proliferation is a challenging duty, as we are not able to divide regions in a pure randomized way and evaluate the effects of circulation restrictions. There is an additional complication of mapping behavioral (non-observable) variables (such as the usage of masks, social distancing and the adoption of better hygiene measures, among many others) that may affect mobility and infection levels, generating an omitted bias issue. On the other hand, when the rates of infection are high, people tend to comply more with restrictive measures, generating a simultaneity bias.

This paper has the objective of analyzing empirically the effects of restrictions to circulation on the infection spread, without recurring to epidemiological models or theoretical formulation of individual behavior. The literature that approached the mobility problem has focused mainly on two points: (i) prediction of the impact of lockdown policies; (ii) evaluation of mobility restrictions in terms of Covid-19 cases and deaths. There is also a third branch that focus on analyzing the effects of Covid-19 on mobility (i.e. the converse of our causal identification), and we only provide some references about such studies.

The first ``prediction" group focus in the impact of lockdown policies in terms of evaluating Covid-19 spread. Inside this group, we also make two distinctions: (i) synthetic controls; (ii) alternative sources of data and models. On the synthetic control subgroup, we point out \cite{carneiro2020lockdown} who adopted an Artificial Counterfactual (ArCo) approach to assess the impacts of the short-run evolution of number of cases (and deaths) in the US. The prediction suggests that, in absence of the restriction measures, the number of cases would be two times larger than observed.
On the same line of using synthetic controls, we point out \cite{bayat2020synthetic}, recurring to a synthetic control methodology to analyze the effects of lockdown measures and the potential impact of those policies on the development of the herd immunity.

On the other subgroup, we focus on studies that used either machine learning models to assess the non-linearities intrinsic to the projection problem (as in \cite{said2020deep}) or the ones who have used alternative sources of data, as Google Mobility (\cite{gerlee2021predicting}) or large-scale mobility data from telecommunication providers (\cite{schwabe2021predicting} and \cite{vespe2021mobility}). These last papers are somehow related to the alternative data sources that we have adopted for controlling the behavioral channel that is not directly measured by conventional variables.

The second group constitutes a larger share of the empirical work and is based on different methodologies to assess the effects of mobility restrictions directly on the evolution of the infection. Based on the availability of Covid-19 infection data, the models are predominantly analyzed in a panel of weekly cases and deaths (to reduce noise effects on daily published data) between or within countries. In terms of methodology, \cite{liu2021panel} suggest the usage of dynamic panel data model to generate forecasts for panel data to capture the inertial elements that affect the infection situation. The authors opt to model the growth rate of the infections, assuming that this variable can be represented by fluctuations around a downward sloping deterministic trend (with a break).

This is also the case of \cite{huang2020effective}, where the author also makes use of the growth rate modelling based on counterfactual analysis to find that social distancing intervention is effective in reducing the weekly growth rate by 9.8\% and deaths by 7.0\% at state-level in the United States. Another example of panel estimation is \cite{chen2020works}
based on a cross-country panel analysis to evaluate each non-pharmaceutical intervention in terms of reducing the reproduction number.
In terms of Brazilian data, \cite{resende2021social} explored a panel-data regression for São Paulo municipalities using labor market dynamics, medical infrastructure and government transfers as controls. The authors found that an increase in 1\% on social distancing reduces infections in 4.14\% in a week and diminishes 2.8\% deaths after two-weeks.

Also in terms of country specific effects of mobility restrictions, \cite{vespe2021mobility} evaluate the effects of restriction on mobility in Italy using mobile network operator data and electricity consumption data to assess the impacts of the Covid-19 wave on the ``three-tier" system. Similarly, \cite{barboza2021role} aim to infer the effects of changes on mobility on the dynamics of the transmission of the Covid-19 in Costa Rica while using Google Mobility to evaluate the effects of sanitary measures.

There are also some other studies that aim to analyze the effects of mobility due to Covid-19 as \cite{engle2020staying}, the effects of non-pharmaceutical interventions in terms of Covid-19 spread as \cite{kong2020disentangling} and also some country specific analyzes which focus on evaluating the effects on mobility after Covid-19 as \cite{batty2021london} for London, \cite{janiak2021covid} for Chile and \cite{benitez2020responses} for Latin-America Countries.

The approach that we have adopted is inserted on the second causal category with some data elements of the prediction group, as we aim to identify how mobility (even in absence of a strict lockdown) affected the infection evolution by recurring to a causal relation framework. We focus on modeling the growth rate of Covid-19 cases and deaths as in \cite{liu2021panel} and \cite{huang2020effective}. Our results are in line with \cite{resende2021social}, but there are three main distinctions in our approach: (i) we model growth rates instead of total number of infections (this avoids the non-stationarity in the infection series due to the high inertial behavior of Covid-19 evolution); (ii) we use national data instead of focusing on a single state analysis; (iii) we adopted soft-data variables to control for non-observable behavioral actions.

In terms of our modelling approach, we created a weekly based panel data for all Brazilian municipalities in order to evaluate the effects of restrictions in mobility in terms of effectiveness while affecting the pandemic evolution. Our paper contributes with the ongoing literature of causal identification of mobility effects based on panel-data evaluation by adding unstructured data (Google Trends and News-indexes) in order to generate proxies for non-observable behavioral variables that affects the Covid-19 spread.

We considered a sample that comprehends the period of May, 2020 - August, 2021 (significantly wider than the studies that focused on the effects of mobility) and we also conduct a sub-sample analysis to capture only non-vaccination periods to evaluate locally the effects of restrictions to circulation. Estimation results suggest that increasing residential mobility (reducing overall mobility) diminishes significantly the number of cases (from 6.19\% on the first week reaching a 3.02\% reduction in four-weeks) and deaths (reducing 2.47\% in one-week growing to a 6.51\% effect in terms of overall reduction in deaths). For the sub-sample period (2020 only) the effects of reducing mobility are similar to the complete sample analysis, but the effects are accumulated with the effect of workplace mobility in terms of cases (deaths): increasing workplace mobility results in a increase in both cases (about 1\%) and deaths (about 2\%) over the reference horizon. The results have been shown to be robust to variations in terms of mobility variables added on the model, geographical aggregation of cases, vaccination campaign variables and Dynamic Panel specifications.

The remainder of this paper is structured in four additional sections. The next section describes the identification strategy, the Direct Acyclic Graphs (DAG) approach, the fixed-effects and dynamic-panel models and all unstructured data that has been created to proxy for non-observable behavioral effects. The third section describes the data. The fourth section describes the estimation results for both all-sample period (2020 and 2021) and only for non-vaccination period (2020 sub-sample). The last section concludes this paper.

\section{Identification Strategy} \label{sec:ident}

We motivate our identification strategy recurring to a Direct Acyclic Graph (DAG) approach, following \cite{elwert2013graphical}. In Appendix A we provide an introduction to the concept of DAGs. In our specification, we want to estimate the impact of mobility on cases (deaths) due to Covid-19, represented by the $\boldsymbol{\beta}$ coefficient. However, there are many confounding factors that may affect mobility (or even the infection situation) that generates an omitted variable bias problem. To overcome such an issue, we specify carefully some of those factors following our hypothesis regarding the causal relation between the variables.

Some prevention measures such as using masks, washing hands and using hand sanitizer may affect the virus infection, i.e. through individual behavior. However, such variable (denoted B, from now on) is non-observable. Therefore, we should include some control variables to capture some of this effect. We use Google Trends searches (gt-series) and News (n-index), both regarding Covid-19 prevention behavior, in order to capture this omitted effect, represented by coefficients $\boldsymbol{\gamma_b}$ and $\boldsymbol{\eta_b}$, respectively. We also include lagged Covid-19 cases (deaths) to capture the lagged effects through tge behavioral channel, inducing a lag structure between Covid-19 spreading over time. In Figure \ref{fig:dag2021} we plot the DAG representing the causal relation between the variables that we adopt in order to identify our model.

We then structure the channels that are represented on the DAG of Figure \ref{fig:dag2021} on the following manner: we consider that vaccination may affect number of cases (deaths), mobility and individual behavior (measured through Google Trends and News proxies with respect to Covid-19 related keywords and searches). We also consider that gt-series and n-index should affect only mobility and may not affect the spread of the disease through direct channels. Therefore, we should have that mobility is mainly determined by vaccination and individual behavior (measured by gt-series and n-index). Nonetheless, we consider that Covid-19 spread is determined by mobility, individual behavior and vaccination.

However, if we analyze solely the first year of the pandemic (2020), the vaccination variable turns out to be innocuous (as the vaccination only started in Brazil by 2021). Therefore, all the channels that relate vaccination with mobility measures, Google Trends searches and Covid-19 related news disappears from the DAG. The result is the DAG present in Figure \ref{fig:dag2020}. Note that the identification of such model is mainly a reduced form of the overall sample model.

After motivating the causal relation that we aim to identify, the econometric specification adopted is straightforward. As we deal with a panel of municipalities at weekly base, we recur to the following functional form:

\begin{equation}
     \ln \left(\frac{Y_{j,t}}{Y_{j,t-1}} \right) \equiv \Delta \ln(Y_{j,t}) = \beta_0 + \boldsymbol{X_{j,t-m}^{\prime}}\boldsymbol{\beta} + \boldsymbol{Z_{j,t}} + \alpha_j + \delta_t + u_{j,t}
    \label{eq:eqn1}
\end{equation}
\noindent where:
\begin{equation*}
    \boldsymbol{Z_{j,t}} = \displaystyle\sum_{h = 1}^{4}\phi_h\ln \left(\frac{Y_{j,t-h}}{Y_{j,t-h-1}}\right) +  \boldsymbol{GT_{j, t-m}^{\prime}}\boldsymbol{\gamma} + \boldsymbol{N_{j,t-m}^{\prime}}\boldsymbol{\eta} + \boldsymbol{V_{j,t-m}}\boldsymbol{\nu}
\end{equation*}

\noindent where $\beta_0, \boldsymbol{\beta}, \boldsymbol{\phi}, \boldsymbol{\gamma}, \boldsymbol{\eta}, \boldsymbol{\nu}, \alpha_j \text{ and } \delta_t $ are parameters to be estimated. Note that $\boldsymbol{\gamma} = (\boldsymbol{\gamma_g}, \boldsymbol{\gamma_b})$ and $\boldsymbol{\eta} = (\boldsymbol{\eta_g}, \boldsymbol{\eta_b)}$ representing gt-series and n-index channels through general Covid-19 related searches or news (g) and through proxies for non-observable behavioral effects (b), respectively. The indexes $j \in \{1, \cdots, 5570\}$ represents municipalities, $t \in \{1, \cdots, 67\}$ is a weekly time-index and $m \in \{1,2,3,4\}$ is the lag structure that we impose over the regressors of the model\footnote{The lag structure is important to capture delayed effects of model's variables in terms of Covid-19 spreading.}. Also, $\alpha_j$ is an individual fixed-effect whereas $\delta_t$ is a time fixed-effect. The constant $\beta_0$ represents a common trend trajectory for all municipalities\footnote{All estimations include an intercept due to the normalization adopted by Stata. Instead of setting the intercept equal to zero, the program adopts the following normalization:
\begin{equation*}
    \sum_{i = 1}^{N} \sum_{t=t_0}^{T} (\beta_0 + \alpha_i) = 0
\end{equation*}}.

In terms of estimation, we impose an within transformation with respect to each municipality and we include time-dummies to capture heterogeneous time-effects throughout time. As robustness check, We also estimate the model using a Dynamic Panel structure (Arellano-Bond transformation with four lags) to capture the effects induced by the inclusion of lagged dependent variables in terms of lag structure, considering the inertial behavioral spreading channel.

Finally, the coefficients associated with mobility and vaccination represents elasticities, i.e. relates the effects of a one percentage change on mobility with the effects over the growth rate of Covid-19 cases (deaths). The coefficients associated with gt-series and the n-index represents semi-elasticities, as they relates the effects of one additional search (or news) with its associated growth-rate effects on Covid-19 spread.

\section{Data} \label{sec:data}

\par There are mainly two types of data (labeled as ``hard data" and ``soft data" from now on) that have been used to estimate our model. The main distinction between them is that the first set is disposed in an objective/highly organized format, whereas the second type relates subjective data that could (e.g. Google Trends) or not (e.g. News-Index) have been organized earlier. Soft data relies on text-related data (unstructured) based on counting measures (structured).

\begin{figure}[H]
\caption{First Symptom and Obit Date versus Notification Date}
    \centering
    \includegraphics[width=16cm]{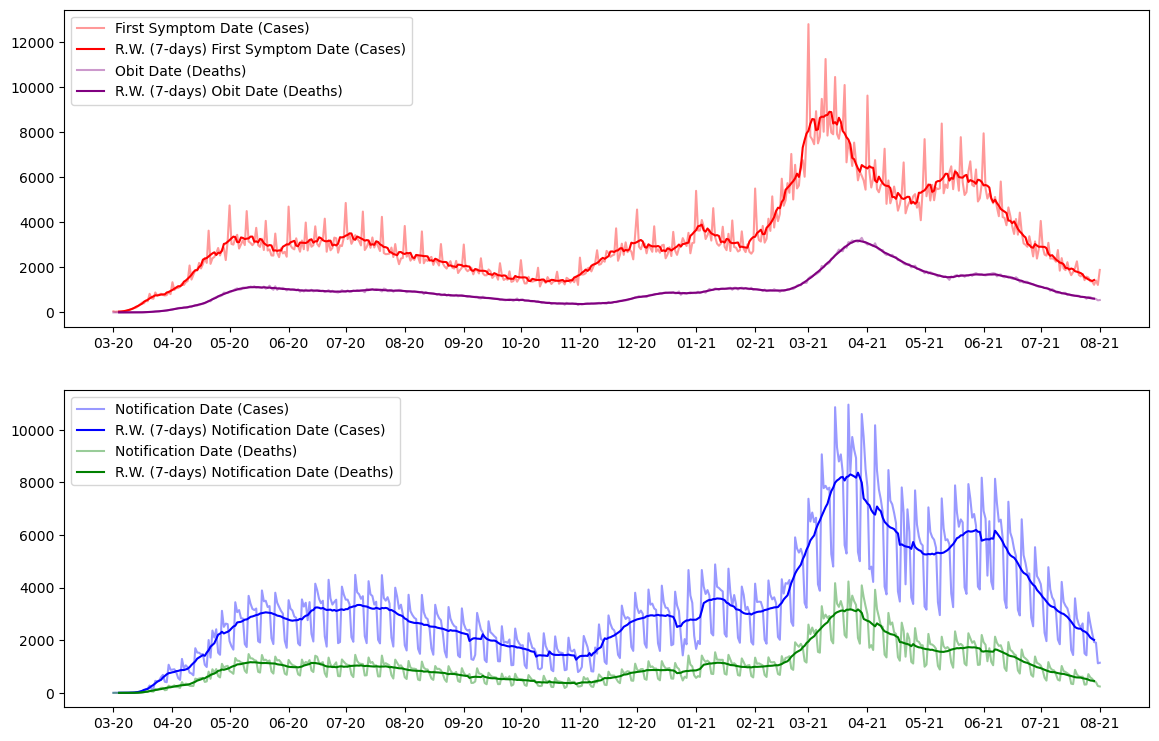}
    \caption*{\footnotesize Comparison between first symptom date (cases) or obit date (deaths) and the notification date at daily basis. The notification date is much more volatile than the effective series for cases (deaths). Regarding the obit date, the series almost coincides with its associated 7-days rolling-window.}
    \label{fig:comp}
\end{figure}

The first set of hard indicators is the number of cases and deaths by Covid-19, which constitutes our dependent variables. Those have been extracted from SRAG data\footnote{SRAG is the acronym for ``Vigilância de Síndrome Respiratória Aguda Grave", which consolidates all data related to harsh respiratory syndrome in Brazil, including Covid-19, for 2020 and 2021. Note that SRAG data only consider patients that effectively enter on hospitals due to a respiratory syndrome and therefore does not represents mild cases.} disposable at the OpenDataSUS website\footnote{OpenDataSUS is an initiative of the Ministry of Health of Brazil.}. The main difference between the construction of Covid-19 cases and deaths series regards the filtering date: for number of cases we have set the first symptom date as reference for aggregation, whereas for number of deaths we set the obit date as reference. In both cases we construct series at the municipality level based on the residence and notification area of the Covid-19 cases (deaths). Such distinction produces different aggregations as they constitute different hypothesis in terms of disease contamination process. In Figure \ref{fig:comp} we compare the effects of different types of date aggregation for number of cases and deaths.

The next set of hard indicators comprehends mobility measures that constitute the object of interest in our estimations. These variables are break down in six mobility categories (workplace, residential, parks, transit, grocery and retail)\footnote{Residence mobility is measured in time spent in-locus, whereas the other five categories are measured in terms of number of visitors.} and compared to the 5‑week baseline period of Jan 3 – Feb 6, 2020, in terms of percentage change. We extract mobility data from the Covid-19 Community Mobility Reports from Google.

Regarding controls, we start by extracting vaccination hard-data regarding timing and immunization type (first or second dose) also from the OpenDataSUS website, which comprehends data at the individual level. As a robustness check, we also collect vaccination from the SRAG data set. The two data sets differs as the first considers the national vaccination campaign, whereas the second considers individuals that are actually inside the SRAG accounting. Both types of vaccination variables are consolidated into a weekly-based period at the municipality level.

The other two variables, Google Trends and News-Index (gt-series and n-index), constitutes our set of soft-controls. Google Trends data have been extracted using the Google Trends API and reveals the number of searches of a given topic for a certain period. News-Index data has been generated based on news collected from G1\footnote{G1 is a local newspaper that belongs to Grupo Globo.} that possess an in-depth coverage of Covid-19 in Brazil\footnote{Only Covid-19 related news constitutes about 142,697 for the period of May 1, 2020 - August 1, 2021. The advantage of considering G1 news is that they are divided into sub-regions, i.e. news-data is locally stamped at the state level.}. These two set of controls relies on a subjective categorization of search terms and keywords selection that needs to be specified in order to generate data-series for our estimates. The formulation of the indexes and the categories/keywords are presented in Appendix B.

Finally, we constitute a weekly base panel data at Municipality level\footnote{Only soft data controls as Google Trends and our News-Index are disposed at State Level, whereas all other variables are disposed in more granular Municipality level} starting on May 3, 2020 up to August 1, 2021\footnote{Two important points: (i) We start (May, 2020) and end our reference panel (August, 2021) to avoid issues regarding both tail data problems or post-publication of statistics from the government, i.e. lack of update of vintages; (ii) the panel data covers a full-week period starting by Monday of a reference week ending at Sunday of the same week.}. Table \ref{tab:stats} and \ref{tab:corr} includes the descriptive statistics and correlation matrix, respectively, for all series used on our estimate.

\section{Results} \label{sec:result}

The results of the estimations are divided into two different sections. The first sub-section analyses the effects of mobility over Covid-19 cases (deaths) through all sample (2020-2021), whereas the second sub-section restricts our model to the sub-sample period of 2020. In both cases, restrictions in mobility tends to display relevant impact on the evolution of the Covid-19 infection rate.

To estimate Equation (\ref{eq:eqn1}), we need to make a consideration regarding the number of lags $m \in \{1,2,3,4\}$ used on our specification. In terms of Covid-19 cases, it is known that the average number of days taken up to the first symptom is about five days (see \cite{cintra2020estimative}). Therefore, inside the 5-days window, the individual may spread the virus without knowing about his infection situation. In terms of deaths, we computed the median number of days taken from the first symptom to obit, represented in Figure \ref{fig:death}. The results suggest that, in the SRAG sample, the median is about seventeen days (third week), while the minimum is about eight days (second week) and the maximum is twenty two days (fourth week). As we consider weekly windows, the horizon that we may consider for cases is from one to four weeks (a month) after the infection and two up to four weeks (a month) for deaths.

Regarding mobility, Table \ref{tab:stats} reveals that only workplace mobility is present in all municipalities that are available in Google's Mobility website. Coming after workplace, residential mobility is also present in a high share overall municipalities. This is not the case for the the other four mobility measures. This generates a higher probability of measurement errors and missing observations in a non-random form. We therefore choose to run the regressions only on the first two measures to obtain a cleaner specification. We also present a full model with all six mobility measures and the results that we find are robust to this inclusion.

\subsection{Complete Sample Estimation: 2020 and 2021}

The estimation results are displayed on Table \ref{tab:res2021}, based on fixed-effects estimations for Covid-19 cases (columns 1 to 4) and deaths (columns 5 to 8). In each column, we highlight the number of lags $m$ growing from $m = 1$ up to $m = 4$ weeks for cases and from $m=2$ to $m=4$ for deaths. The coefficient of interest is the residential and workplace mobility, as they are associated with the highest number of observations for municipalities. The first important result is that increasing residential mobility (i.e. decreasing overall mobility) reduces the growth rate of both number of infections and deaths through the sample. The impact over Covid-19 cases is higher over the first reference week, where it diminishes cases at a 6.19\% rate, and slowly decreases through the 4-week window, reaching 3.02\% after one month. Regarding Covid-19 deaths, we also observe that reducing mobility affects negatively the growth rate of deaths. However, this effect grows from 2.47\% reaching 6.51\% at a month horizon. If we consider that this effect can be combined week-by-week (inducing an over-estimate of the overall effect) we can determine that the overall (maximum) upper bound effect of a 1\% decrease in mobility results in a reduction of 20.83\% in cases and a reduction in 14.35\% in deaths, both at a one-month horizon.

The vaccination, gt-series and n-index controls displays important role while capturing the effects suggested on the DAGs. This is also observable with the inertial behavior that seems to be captured on the coefficients associated with the lagged dependent variables. The overall R-squared is at least 18\% for cases and 23\% for deaths (higher explanatory power), and the F-test rejects the null hypothesis that all coefficients are zero for all estimates. All lagged dependent variables display negative inertial effect on the evolution of both cases and deaths.

On Appendix C we display first stage estimations with the objective of validating the DAG channels presented on Figure \ref{fig:dag2021} assumptions displayed on Section \ref{sec:ident}. In Table \ref{tab:1stage} we display estimation results of mobility against gt-series, n-index and lagged dependent variables and we find high associated R-squared and null rejection of F-test. Table \ref{tab:vac/cases} relates cases (deaths) regression against vaccination campaign, finding mostly negative coefficients as expected. Table \ref{tab:vac/mob} relates mobility with the national vaccination campaign and with SRAG vaccination, finding high associated R-squared. It is important to notice that the SRAG data has higher explanatory power, but only consider vaccination on individuals inside the SRAG data set, which is restrictive in terms of causal determination.

We also display a regression with all six mobility measures on Appendix D. The results present in Table \ref{tab:moball} suggests some ambiguity: transit and grocery seems to display a negative impact on the growth rate of cases and deaths. We do not take these estimated coefficients as  pure effects due to: (i) lack of observations in smaller counties; (ii) the effects are smaller than 0.5\%, to little to be tacking into account. Additionally, Appendix D display three different estimation outputs. Table \ref{tab:not2021} uses notification area instead of residence place while aggregating Covid-19 cases (deaths). The results are similar to the ones that has been founded using residence area. Table \ref{tab:vac} uses SRAG vaccination data (log of vaccines for SRAG) instead of national campaign vaccination data (that comprehends first and second dose breakdown), with similar results. Finally, Table \ref{tab:dynpan} uses a Dynamic-Panel (Arellano-Bond with four lags) methodology (in line with \cite{liu2021panel}) to estimate recursively the effects of the lagged variable impact over the dependent variable and the residential mobility estimates still reveals a negative sign and with decaying (increasing) effects for cases (deaths).

\subsection{Sub-sample Estimation: 2020}

This section has the objective of estimating the impact of mobility on Covid-19 cases (deaths) only at the first year of the Covid-19 pandemic. By limiting the sample solely for 2020, we can redefine the causal relation that we aim to identify as the vaccination only started in 2021. This results on the DAG that is presented on Figure \ref{fig:dag2020}, in which we remove the vaccination node and solely focus on mobility-related variables.

The estimation results are presented on Table \ref{tab:res2020} and two effects are directly observable: (i) the negative sign associated with residential mobility is still present; (ii) workplace mobility displays an important role while explaining number of cases (deaths) due to Covid-19 spread. The magnitude of the effects of reducing mobility and growth rate of number of cases is still on the order of 5.04\% on a first-week horizon, decreasing to 4.08\% at a month. For deaths, this effect frows from 3.31\% over two weeks up to 6.25\% after a month. The effects of increasing mobility, however, tends to increase over the estimation window for both cases and deaths growth rate, oscillating positively around 1-2\%.

By recurring to the same exercise of combining the effects in order to generate an upper bound for mobility effects, we can check that a 1\% mobility increase for workplace generates a similar 20.73\% and 16.36\% reduction for cases and deaths, respectively. There is however the effect of the reduction on workplace mobility, that a 1\% decrease can combined generate a 4.8\% and 6.07\% decrease in cases and deaths, respectively. In Appendix D we also display Dynamic-Panel estimates for the 2020 sub-sample and the coefficients still presents compatible magnitudes to the ones observed in Table \ref{tab:res2020}.

\section{Conclusion} \label{sec:conclusion}

In this paper we aimed in developing an empirical framework to be able to address the questions related to causal effects of mobility restrictions in terms of Covid-19 infections. By recurring to a DAG approach, we developed causal channels that required the design of ``soft-data" proxies to capture non-observable effects of individuals, as prevention measures.

The methodology adopted to estimate the effects is in line with the panel-data estimates by \cite{liu2021panel} and \cite{huang2020effective} with even similar dimensions in terms of the mobility elasticity sizes: a 1\% reduction in mobility (through an increase in residential mobility) reduces cases and deaths in a one-month horizon for both 2020-2021 sample and 2020-only. The combined upper bound effect is around a 20\% reduction for cases and 15\% reduction for deaths.

There are some limitations in our analysis based on potential non-linearities on the effect of vaccination over mobility: vaccines can induce higher or lower mobility depending on the overall vaccination campaign or even on the Covid-19 infection level. We therefore avoid interpreting vaccination related coefficients as they can be misleading. The same happens while analyzing other mobility coefficients, due to the intrinsic measurement error in those variables.

Finally, the paper suggests that restricting mobility is able to reduce the number of cases and deaths with particular robustness throughout sample and methodological evaluation. The objective of developing a causal framework based on DAGs is sustained by the estimation outputs and the mobility effects are solely determined by the causal hypothesis.

\newpage
\singlespacing
\setlength\bibsep{0pt}
\bibliographystyle{plainnat}
\bibliography{ref.bib}

\clearpage

\newpage

\section*{Tables} \label{sec:tab}
\addcontentsline{toc}{section}{Tables}

\begin{table}[H]
\centering
\caption{Descriptive Statistics for Model's variables\label{tab:stats}}
\begin{adjustbox}{max width=\textwidth}
\begin{threeparttable}
\begin{tabular}{c|cccccccccccc}
\textbf{Mun}   & \textbf{date}     & \textbf{cases}         & \textbf{deaths}      & \textbf{residential} & \textbf{workplace} & \textbf{transit}        & \textbf{parks}        & \textbf{grocery}      & \textbf{retail} & \textbf{1st\_dose} & \textbf{2nd\_dose} & \textbf{srag\_vac} \\ \hline
\textbf{count} & 67                & 169598                 & 103767               & 63025                & 142060             & 39528                   & 56603                 & 53758                 & 57023           & 130047             & 111416             & 27866              \\
\textbf{mean}  & -                 & 9.67                   & 5.02                 & 7.93                 & -3.52              & -27.03                  & -31.09                & 13.79                 & -22.87          & 695.88             & 312.41             & 4.19               \\
\textbf{min}   & 2020-05-03        & 1                      & 1                    & -22.14               & -82                & -100                    & -100                  & -91                   & -92.14          & 1                  & 1                  & 1                  \\
\textbf{50\%}  & -                 & 2                      & 1                    & 7.57                 & -2.67              & -31                     & -33                   & 14                    & -21.71          & 144                & 53                 & 1                  \\
\textbf{max}   & 2021-08-01        & 5571                   & 1606                 & 29.2                 & 57.4               & 246.43                  & 313                   & 148                   & 187             & 647665             & 339556             & 544                \\
\textbf{std}   & -                 & 58.96                  & 24.38                & 4.11                 & 12.5               & 31.48                   & 29.45                 & 22.55                 & 21.59           & 5688.22            & 2870.51            & 15.05              \\ \hline
\textbf{UF}    & \textbf{n\_covid} & \textbf{n\_prevention} & \textbf{n\_fakenews} & \textbf{n\_vaccine}  & \textbf{gt\_covid} & \textbf{gt\_prevention} & \textbf{gt\_fakenews} & \textbf{gt\_vaccines} & \textbf{}       & \textbf{}          & \textbf{}          &                    \\ \hline
\textbf{count} & 2025              & 2025                   & 2025                 & 2025                 & 2079               & 2079                    & 2079                  & 2079                  &                 &                    &                    &                    \\
\textbf{mean}  & 43.22667          & 3.595556               & 0.142716             & 9.532346             & 817.6402           & 189.7946                & 92.06686              & 197.8066              &                 &                    &                    &                    \\
\textbf{std}   & 67.58067          & 7.465526               & 0.52117              & 21.37178             & 392.4496           & 117.3915                & 76.17474              & 167.4312              &                 &                    &                    &                    \\
\textbf{min}   & 0                 & 0                      & 0                    & 0                    & 29                 & 0                       & 0                     & 0                     &                 &                    &                    &                    \\
\textbf{50\%}  & 24                & 1                      & 0                    & 2                    & 784                & 164                     & 73                    & 125                   &                 &                    &                    &                    \\
\textbf{max}   & 520               & 95                     & 7                    & 219                  & 2297               & 768                     & 435                   & 829                   &                 &                    &                    &                    \\ \hline
\end{tabular}
\begin{tablenotes}
\small
\item[a] The descriptive statistics table reveals the statistics in relation of municipalities level variables (5570 counties) and federal unity variables (27 states) throughout 67-weeks.
\end{tablenotes}
\end{threeparttable}
\end{adjustbox}
\end{table}


\bigskip

\begin{table}[H]
\centering
\caption{Correlation Matrix for Model's variables\label{tab:corr}}
\begin{adjustbox}{totalheight = 8cm, max width=\textwidth}
\begin{threeparttable}
\begin{tabular}{cccccccccccccccccccc}
\multicolumn{1}{c|}{}                     & \textbf{cases} & \textbf{deaths} & \textbf{residential} & \textbf{workplace} & \textbf{transit} & \textbf{parks} & \textbf{grocery} & \textbf{retail} & \textbf{1st\_dose} & \textbf{2nd\_dose} & \textbf{srag\_vac} & \textbf{n\_covid} & \textbf{n\_prevention} & \textbf{n\_fakenews} & \textbf{n\_vaccines} & \textbf{gt\_covid} & \textbf{gt\_prevention} & \textbf{gt\_fakenews} & \textbf{gt\_vaccines} \\ \hline
\multicolumn{1}{c|}{\textbf{cases}}       & 1              &                 &                      &                    &                  &                &                  &                 &                    &                    &                    &                   &                        &                      &                      &                    &                         &                       &                       \\
\multicolumn{1}{c|}{\textbf{deaths}}      & 0.888811       & 1               &                      &                    &                  &                &                  &                 &                    &                    &                    &                   &                        &                      &                      &                    &                         &                       &                       \\
\multicolumn{1}{c|}{\textbf{residential}} & 0.13881        & 0.191111        & 1                    &                    &                  &                &                  &                 &                    &                    &                    &                   &                        &                      &                      &                    &                         &                       &                       \\
\multicolumn{1}{c|}{\textbf{workplace}}   & -0.09965       & -0.13758        & -0.6129              & 1                  &                  &                &                  &                 &                    &                    &                    &                   &                        &                      &                      &                    &                         &                       &                       \\
\multicolumn{1}{c|}{\textbf{transit}}     & 0.012221       & -0.01424        & -0.22639             & 0.288031           & 1                &                &                  &                 &                    &                    &                    &                   &                        &                      &                      &                    &                         &                       &                       \\
\multicolumn{1}{c|}{\textbf{parks}}       & 0.001204       & -0.02411        & -0.20605             & 0.293545           & 0.3163           & 1              &                  &                 &                    &                    &                    &                   &                        &                      &                      &                    &                         &                       &                       \\
\multicolumn{1}{c|}{\textbf{grocery}}     & -0.01777       & -0.04402        & -0.37866             & 0.492935           & 0.417557         & 0.461792       & 1                &                 &                    &                    &                    &                   &                        &                      &                      &                    &                         &                       &                       \\
\multicolumn{1}{c|}{\textbf{retail}}      & -0.09104       & -0.13524        & -0.58645             & 0.655786           & 0.464387         & 0.52301        & 0.666516         & 1               &                    &                    &                    &                   &                        &                      &                      &                    &                         &                       &                       \\
\multicolumn{1}{c|}{\textbf{1st\_dose}}                        & 0.620561       & 0.646347        & 0.142404             & -0.01313           & 0.055703         & 0.071746       & 0.057885         & -0.02395        & 1                  &                    &                    &                   &                        &                      &                      &                    &                         &                       &                       \\
\multicolumn{1}{c|}{\textbf{2nd\_dose}}                        & 0.534721       & 0.550117        & 0.120569             & -0.02577           & 0.036762         & 0.050639       & 0.028651         & -0.03392        & 0.688053           & 1                  &                    &                   &                        &                      &                      &                    &                         &                       &                       \\
\multicolumn{1}{c|}{\textbf{srag\_vac}}                        & 0.701067       & 0.717479        & 0.172842             & -0.06906           & 0.014043         & 0.057704       & 0.000136         & -0.08699        & 0.810688           & 0.659897           & 1                  &                   &                        &                      &                      &                    &                         &                       &                       \\
\multicolumn{1}{c|}{\textbf{n\_covid}}                         & 0.064728       & 0.064558        & 0.168983             & -0.11722           & -0.06797         & 0.062322       & -0.03967         & -0.18957        & 0.043605           & 0.028973           & 0.073001           & 1                 &                        &                      &                      &                    &                         &                       &                       \\
\multicolumn{1}{c|}{\textbf{n\_prevention}}                    & 0.058169       & 0.068981        & 0.348067             & -0.29392           & -0.15638         & -0.02013       & -0.19161         & -0.3224         & 0.026125           & 0.006129           & 0.055414           & 0.754311          & 1                      &                      &                      &                    &                         &                       &                       \\
\multicolumn{1}{c|}{\textbf{n\_fakenews}}                      & 0.031402       & 0.040314        & 0.137724             & -0.11682           & -0.06677         & -0.00962       & -0.07884         & -0.1635         & 0.008005           & 0.010872           & 0.01946            & 0.477066          & 0.432017               & 1                    &                      &                    &                         &                       &                       \\
\multicolumn{1}{c|}{\textbf{n\_vaccines}}                      & 0.051911       & 0.046724        & -0.14051             & 0.09323            & 0.054884         & 0.035115       & 0.179795         & 0.043606        & 0.063243           & 0.041721           & 0.085583           & 0.628181          & 0.237822               & 0.24393              & 1                    &                    &                         &                       &                       \\
\multicolumn{1}{c|}{\textbf{gt\_covid}}                        & 0.059832       & 0.069618        & 0.079174             & -0.15763           & -0.09175         & -0.13204       & -0.01294         & -0.23743        & 0.021661           & -0.01218           & 0.040347           & 0.440123          & 0.361777               & 0.204884             & 0.480377             & 1                  &                         &                       &                       \\
\multicolumn{1}{c|}{\textbf{gt\_prevention}}                   & 0.048256       & 0.051827        & 0.333856             & -0.35703           & -0.21313         & -0.15469       & -0.28985         & -0.38402        & -0.01588           & -0.02455           & -0.0019            & 0.350247          & 0.491897               & 0.212462             & 0.101948             & 0.59389            & 1                       &                       &                       \\
\multicolumn{1}{c|}{\textbf{gt\_fakenews}}                     & 0.049763       & 0.067476        & 0.256313             & -0.29008           & -0.17234         & -0.16648       & -0.17583         & -0.34397        & -0.00736           & -0.02306           & 0.011633           & 0.277374          & 0.35914                & 0.178029             & 0.159533             & 0.704878           & 0.510057                & 1                     &                       \\
\multicolumn{1}{c|}{\textbf{gt\_vaccines}}                     & 0.018039       & 0.020033        & -0.26664             & 0.224302           & 0.131146         & 0.021521       & 0.319254         & 0.161632        & 0.064489           & 0.022439           & 0.04706            & 0.179439          & -0.0397                & 0.028723             & 0.551505             & 0.65084            & -0.02194                & 0.149079              & 1             \\
\hline
\end{tabular}
\begin{tablenotes}
\Large
\item[a] Correlation matrix for all variables and series used on the main estimate of the article. Note that mobility measures display internal consistence: by construction, residence mobility is negatively correlated to all non-residential mobility series.
\end{tablenotes}
\end{threeparttable}
\end{adjustbox}
\end{table}

\newpage

\bigskip

\begin{table}
\centering
\caption{Estimation for Complete Sample (2020-2021) by Residence Area\label{tab:res2021}}
\begin{adjustbox}{max width= 13cm}
\begin{threeparttable}
{
\def\sym#1{\ifmmode^{#1}\else\(^{#1}\)\fi}
\begin{tabular}{l*{7}{c}}
\hline\hline
            &\multicolumn{1}{c}{(1)}&\multicolumn{1}{c}{(2)}&\multicolumn{1}{c}{(3)}&\multicolumn{1}{c}{(4)}&\multicolumn{1}{c}{(5)}&\multicolumn{1}{c}{(6)}&\multicolumn{1}{c}{(7)}\\
            &\multicolumn{1}{c}{cases}&\multicolumn{1}{c}{cases}&\multicolumn{1}{c}{cases}&\multicolumn{1}{c}{cases}&\multicolumn{1}{c}{deaths}&\multicolumn{1}{c}{deaths}&\multicolumn{1}{c}{deaths}\\
            &         m=1         &         m=2         &         m=3         &         m=4         &         m=2         &         m=3         &         m=4         \\
\hline
residential &     -0.0619\sym{***}&     -0.0565\sym{***}&     -0.0455\sym{***}&     -0.0302\sym{***}&     -0.0247\sym{***}&     -0.0478\sym{***}&     -0.0651\sym{***}\\
            &   (0.00370)         &   (0.00407)         &   (0.00432)         &   (0.00397)         &   (0.00469)         &   (0.00482)         &   (0.00643)         \\
[1em]
workplace   &    0.000115         &  -0.0000540         &    0.000886         &    0.000787         &     0.00173         &     0.00328\sym{*}  &     0.00260         \\
            &   (0.00104)         &   (0.00113)         &   (0.00116)         &   (0.00114)         &   (0.00147)         &   (0.00161)         &   (0.00199)         \\
[1em]
1st\_dose    &      0.0104\sym{*}  &    -0.00200         &    -0.00523         &    -0.00402         &     0.00421         &    0.000462         &    -0.00445         \\
            &   (0.00452)         &   (0.00476)         &   (0.00478)         &   (0.00502)         &   (0.00695)         &   (0.00676)         &   (0.00718)         \\
[1em]
2nd\_dose    &     0.00312         &    -0.00686\sym{*}  &    -0.00378         &    -0.00571         &   -0.000302         &    -0.00771         &    -0.00452         \\
            &   (0.00318)         &   (0.00300)         &   (0.00306)         &   (0.00331)         &   (0.00450)         &   (0.00465)         &   (0.00447)         \\
[1em]
n\_covid     &    0.000605\sym{***}&    0.000462\sym{**} &    0.000368\sym{*}  &    0.000654\sym{***}&    0.000233         &    0.000229         &   0.0000937         \\
            &  (0.000165)         &  (0.000164)         &  (0.000174)         &  (0.000163)         &  (0.000272)         &  (0.000256)         &  (0.000246)         \\
[1em]
n\_prevention&    0.000590         &    0.000274         &    0.000346         &   -0.000329         &    -0.00236\sym{*}  &    0.000739         &     0.00195         \\
            &  (0.000714)         &  (0.000763)         &  (0.000720)         &  (0.000819)         &   (0.00114)         &   (0.00109)         &   (0.00109)         \\
[1em]
n\_fakenews  &     -0.0172\sym{***}&     -0.0265\sym{***}&   -0.000623         &    -0.00133         &     -0.0132         &     -0.0177\sym{**} &     -0.0196\sym{**} \\
            &   (0.00462)         &   (0.00476)         &   (0.00473)         &   (0.00467)         &   (0.00695)         &   (0.00686)         &   (0.00658)         \\
[1em]
n\_vaccines  &   -0.000262         &    -0.00107\sym{***}&    -0.00104\sym{***}&    -0.00125\sym{***}&    0.000629         &   -0.000127         &   -0.000353         \\
            &  (0.000260)         &  (0.000270)         &  (0.000295)         &  (0.000320)         &  (0.000360)         &  (0.000383)         &  (0.000435)         \\
[1em]
gt\_covid    &   0.0000713         &   -0.000145\sym{***}&   -0.000381\sym{***}&   -0.000491\sym{***}&    0.000453\sym{***}&    0.000193\sym{**} &    0.000153\sym{*}  \\
            & (0.0000419)         & (0.0000431)         & (0.0000440)         & (0.0000455)         & (0.0000685)         & (0.0000636)         & (0.0000630)         \\
[1em]
gt\_prevention&   0.0000349         &    0.000165         &    0.000120         &  -0.0000108         &   0.0000275         &   0.0000181         &   -0.000192         \\
            & (0.0000914)         & (0.0000948)         & (0.0000923)         & (0.0000941)         &  (0.000140)         &  (0.000132)         &  (0.000131)         \\
[1em]
gt\_fakenews &   0.0000524         &   -0.000223         &    0.000116         &    0.000210         &    0.000172         &    0.000314         &   -0.000426\sym{*}  \\
            &  (0.000123)         &  (0.000127)         &  (0.000132)         &  (0.000125)         &  (0.000186)         &  (0.000182)         &  (0.000186)         \\
[1em]
gt\_vaccines &   -0.000302\sym{***}&  -0.0000752         &    0.000192\sym{*}  &    0.000488\sym{***}&   -0.000735\sym{***}&   -0.000336\sym{**} &   -0.000511\sym{***}\\
            & (0.0000824)         & (0.0000808)         & (0.0000844)         & (0.0000875)         &  (0.000115)         &  (0.000121)         &  (0.000132)         \\
[1em]
cases\_1  &      -0.492\sym{***}&      -0.504\sym{***}&      -0.505\sym{***}&      -0.510\sym{***}&                     &                     &                     \\
            &   (0.00865)         &   (0.00891)         &   (0.00918)         &   (0.00939)         &                     &                     &                     \\
[1em]
cases\_2  &      -0.242\sym{***}&      -0.263\sym{***}&      -0.274\sym{***}&      -0.274\sym{***}&                     &                     &                     \\
            &   (0.00949)         &   (0.00971)         &   (0.00979)         &    (0.0102)         &                     &                     &                     \\
[1em]
cases\_3  &      -0.118\sym{***}&      -0.136\sym{***}&      -0.150\sym{***}&      -0.154\sym{***}&                     &                     &                     \\
            &   (0.00940)         &   (0.00922)         &   (0.00946)         &   (0.00981)         &                     &                     &                     \\
[1em]
cases\_4  &     -0.0408\sym{***}&     -0.0617\sym{***}&     -0.0708\sym{***}&     -0.0741\sym{***}&                     &                     &                     \\
            &   (0.00798)         &   (0.00806)         &   (0.00819)         &   (0.00842)         &                     &                     &                     \\
[1em]
deaths\_1 &                     &                     &                     &                     &      -0.602\sym{***}&      -0.611\sym{***}&      -0.628\sym{***}\\
            &                     &                     &                     &                     &    (0.0107)         &    (0.0105)         &    (0.0109)         \\
[1em]
deaths\_2 &                     &                     &                     &                     &      -0.337\sym{***}&      -0.345\sym{***}&      -0.371\sym{***}\\
            &                     &                     &                     &                     &    (0.0129)         &    (0.0125)         &    (0.0125)         \\
[1em]
deaths\_3 &                     &                     &                     &                     &      -0.157\sym{***}&      -0.160\sym{***}&      -0.184\sym{***}\\
            &                     &                     &                     &                     &    (0.0115)         &    (0.0114)         &    (0.0111)         \\
[1em]
deaths\_4 &                     &                     &                     &                     &     -0.0624\sym{***}&     -0.0567\sym{***}&     -0.0676\sym{***}\\
            &                     &                     &                     &                     &   (0.00971)         &   (0.00988)         &   (0.00959)         \\
[1em]
\_cons      &       1.456\sym{***}&       1.344\sym{***}&       1.450\sym{***}&       0.961\sym{***}&       1.307\sym{***}&       1.363\sym{***}&       1.622\sym{***}\\
            &    (0.0567)         &    (0.0614)         &    (0.0652)         &    (0.0627)         &    (0.0867)         &    (0.0833)         &    (0.0914)         \\
\hline
$R^2$          &       0.265         &       0.263         &       0.262         &       0.254         &       0.302         &       0.307         &       0.318         \\
$R_{overall}^2$        &       0.186         &       0.181         &       0.183         &       0.199         &       0.262         &       0.251         &       0.234         \\
N           &       21084         &       20309         &       19516         &       18723         &       14049         &       13645         &       13228         \\
p           &           .         &           .         &           .         &           .         &           .         &           .         &           .         \\
\hline\hline
\multicolumn{8}{l}{\footnotesize Standard errors in parentheses}\\
\multicolumn{8}{l}{\footnotesize \sym{*} \(p<0.05\), \sym{**} \(p<0.01\), \sym{***} \(p<0.001\)}\\
\end{tabular}
}

\begin{tablenotes}
\small
\item[a]{Results for Fixed-Effects Estimation over 2020 and 2021 sample, aggregating cases (deaths) by residence and using overall vaccination from OpenData SUS data set. $R^2$ denotes the R-squared, $R_{overall}^2$ the overall R-squared, $N$ the total number of observations used on the estimation and $p$ the F-test associated p-value.}
\end{tablenotes}
\end{threeparttable}
\end{adjustbox}
\end{table}

\newpage

\bigskip

\begin{table}
\centering
\caption{Estimation for Sub-Sample (2020) by Residence Area\label{tab:res2020}}
\begin{adjustbox}{max width= 14cm}
\begin{threeparttable}
{
\def\sym#1{\ifmmode^{#1}\else\(^{#1}\)\fi}
\begin{tabular}{l*{7}{c}}
\hline\hline
            &\multicolumn{1}{c}{(1)}&\multicolumn{1}{c}{(2)}&\multicolumn{1}{c}{(3)}&\multicolumn{1}{c}{(4)}&\multicolumn{1}{c}{(5)}&\multicolumn{1}{c}{(6)}&\multicolumn{1}{c}{(7)}\\
            &\multicolumn{1}{c}{cases}&\multicolumn{1}{c}{cases}&\multicolumn{1}{c}{cases}&\multicolumn{1}{c}{cases}&\multicolumn{1}{c}{deaths}&\multicolumn{1}{c}{deaths}&\multicolumn{1}{c}{deaths}\\
            &         m=1         &         m=2         &         m=3         &         m=4         &         m=2         &         m=3         &         m=4         \\
\hline
residential &     -0.0504\sym{***}&     -0.0518\sym{***}&     -0.0500\sym{***}&     -0.0408\sym{***}&     -0.0331\sym{***}&     -0.0601\sym{***}&     -0.0625\sym{***}\\
            &   (0.00551)         &   (0.00526)         &   (0.00545)         &   (0.00456)         &   (0.00766)         &   (0.00769)         &   (0.00865)         \\
[1em]
workplace   &      0.0120\sym{***}&      0.0127\sym{***}&      0.0118\sym{***}&      0.0108\sym{***}&      0.0181\sym{***}&      0.0201\sym{***}&      0.0214\sym{***}\\
            &   (0.00170)         &   (0.00154)         &   (0.00149)         &   (0.00145)         &   (0.00269)         &   (0.00272)         &   (0.00303)         \\
[1em]
n\_covid     &   -0.000142         &   -0.000249         &   -0.000305\sym{*}  &   -0.000377\sym{**} &   -0.000155         &    0.000105         &   -0.000224         \\
            &  (0.000147)         &  (0.000139)         &  (0.000131)         &  (0.000130)         &  (0.000278)         &  (0.000267)         &  (0.000249)         \\
[1em]
n\_prevention&   0.0000281         &    0.000180         &     0.00141\sym{*}  &     0.00141\sym{*}  &   -0.000152         &   -0.000886         &    0.000455         \\
            &  (0.000800)         &  (0.000692)         &  (0.000655)         &  (0.000648)         &   (0.00115)         &   (0.00111)         &   (0.00100)         \\
[1em]
n\_fakenews  &    0.000370         &     0.00398         &     0.00208         &     0.00730         &      0.0113         &    -0.00837         &      0.0105         \\
            &   (0.00763)         &   (0.00749)         &   (0.00707)         &   (0.00712)         &    (0.0109)         &    (0.0107)         &    (0.0103)         \\
[1em]
n\_vaccines  &    0.000202         &     0.00162         &   -0.000182         &   -0.000296         &     -0.0105\sym{***}&    -0.00441         &    -0.00171         \\
            &   (0.00196)         &   (0.00191)         &   (0.00199)         &   (0.00174)         &   (0.00282)         &   (0.00302)         &   (0.00268)         \\
[1em]
gt\_covid    &    0.000240\sym{***}&   0.0000696         &   -0.000157\sym{*}  &   -0.000396\sym{***}&    0.000500\sym{***}&    0.000319\sym{**} &    0.000275\sym{**} \\
            & (0.0000643)         & (0.0000657)         & (0.0000680)         & (0.0000638)         & (0.0000958)         &  (0.000101)         &  (0.000104)         \\
[1em]
gt\_prevention&  -0.0000715         &  -0.0000352         &   -0.000194         &    0.000325\sym{**} &   -0.000131         &   -0.000171         &   -0.000564\sym{**} \\
            &  (0.000137)         &  (0.000130)         &  (0.000128)         &  (0.000126)         &  (0.000195)         &  (0.000198)         &  (0.000209)         \\
[1em]
gt\_fakenews &  -0.0000836         &   -0.000322         &   -0.000234         &   -0.000233         &   0.0000561         &   0.0000297         &   -0.000554\sym{*}  \\
            &  (0.000167)         &  (0.000169)         &  (0.000171)         &  (0.000162)         &  (0.000262)         &  (0.000253)         &  (0.000223)         \\
[1em]
gt\_vaccines &  -0.0000730         &   0.0000370         &    0.000390         &   0.0000451         &   -0.000136         &   -0.000334         &   0.0000769         \\
            &  (0.000202)         &  (0.000213)         &  (0.000223)         &  (0.000230)         &  (0.000329)         &  (0.000321)         &  (0.000325)         \\
[1em]
cases\_1  &      -0.500\sym{***}&      -0.499\sym{***}&      -0.505\sym{***}&      -0.509\sym{***}&                     &                     &                     \\
            &   (0.00995)         &   (0.00983)         &   (0.00980)         &    (0.0101)         &                     &                     &                     \\
[1em]
cases\_2  &      -0.228\sym{***}&      -0.236\sym{***}&      -0.244\sym{***}&      -0.246\sym{***}&                     &                     &                     \\
            &    (0.0113)         &    (0.0111)         &    (0.0110)         &    (0.0110)         &                     &                     &                     \\
[1em]
cases\_3  &     -0.0986\sym{***}&      -0.110\sym{***}&      -0.124\sym{***}&      -0.126\sym{***}&                     &                     &                     \\
            &    (0.0100)         &    (0.0100)         &   (0.00967)         &   (0.00947)         &                     &                     &                     \\
[1em]
cases\_4  &     -0.0441\sym{***}&     -0.0474\sym{***}&     -0.0571\sym{***}&     -0.0625\sym{***}&                     &                     &                     \\
            &   (0.00867)         &   (0.00874)         &   (0.00861)         &   (0.00849)         &                     &                     &                     \\
[1em]
deaths\_1 &                     &                     &                     &                     &      -0.606\sym{***}&      -0.618\sym{***}&      -0.632\sym{***}\\
            &                     &                     &                     &                     &    (0.0140)         &    (0.0139)         &    (0.0139)         \\
[1em]
deaths\_2 &                     &                     &                     &                     &      -0.327\sym{***}&      -0.342\sym{***}&      -0.361\sym{***}\\
            &                     &                     &                     &                     &    (0.0164)         &    (0.0162)         &    (0.0163)         \\
[1em]
deaths\_3 &                     &                     &                     &                     &      -0.110\sym{***}&      -0.131\sym{***}&      -0.146\sym{***}\\
            &                     &                     &                     &                     &    (0.0149)         &    (0.0148)         &    (0.0150)         \\
[1em]
deaths\_4 &                     &                     &                     &                     &     -0.0369\sym{**} &     -0.0480\sym{***}&     -0.0623\sym{***}\\
            &                     &                     &                     &                     &    (0.0122)         &    (0.0121)         &    (0.0121)         \\
[1em]
\_cons      &       1.590\sym{***}&       1.816\sym{***}&       2.175\sym{***}&       2.089\sym{***}&       1.647\sym{***}&       2.589\sym{***}&       2.977\sym{***}\\
            &    (0.0945)         &    (0.0892)         &    (0.0953)         &    (0.0920)         &     (0.163)         &     (0.171)         &     (0.162)         \\
\hline
$R^2$          &       0.232         &       0.234         &       0.236         &       0.238         &       0.294         &       0.306         &       0.312         \\
$R_{overall}^2$        &       0.171         &       0.169         &       0.169         &       0.176         &       0.224         &       0.208         &       0.205         \\
N           &       17598         &       17531         &       17478         &       17433         &        8471         &        8468         &        8465         \\
p           &           0         &   1.08e-286         &   4.86e-287         &   2.37e-285         &   4.16e-188         &   2.11e-191         &   2.25e-192         \\
\hline\hline
\multicolumn{8}{l}{\footnotesize Standard errors in parentheses}\\
\multicolumn{8}{l}{\footnotesize \sym{*} \(p<0.05\), \sym{**} \(p<0.01\), \sym{***} \(p<0.001\)}\\
\end{tabular}
}

\begin{tablenotes}
\small
\item[a]{Results for Fixed-Effects Estimation for 2020 only, aggregating cases (deaths) by residence. $R^2$ denotes the R-squared, $R_{overall}^2$ the overall R-squared, $N$ the total number of observations used on the estimation and $p$ the F-test associated p-value.}
\end{tablenotes}
\end{threeparttable}
\end{adjustbox}
\end{table}

\clearpage

\section*{Figures} \label{sec:fig}
\addcontentsline{toc}{section}{Figures}

\begin{figure}[H]
\caption{DAG for 2021 identification representing causal chain within model's variables}
    \label{fig:dag2021}
    \centering
    \begin{tikzpicture}
\node[text centered] (X) {$X$};
\node[below left = 1 of X, text centered] (Ng) {$N_{g}$};
\node[below=0.5 of Ng, text centered] (Gg) {$G_{g}$};
\node[above=1 of X, text centered] (B) {$\mycirc{$B$}$};
\node[above left=0.5 of B, text centered] (Gb) {$G_{b}$};
\node[above=1 of B, text centered] (Ylag) {$Y_{\text{lag}}$};
\node[above right = 0.5 of B, text centered] (Nb) {$N_{b}$};
\node[left = 3 of X, text centered] (V) {$V$};
\node[right = 3 of X, text centered] (Y) {$Y$};

\draw[->, line width= 1] (X) -- node[above,font=\footnotesize]{$\boldsymbol{\beta}$}  (Y);
\draw[->, line width= 1] (V) -- node[above,font=\footnotesize]{}  (X);
\draw[->, line width= 1] (Gg) -- node[above,font=\footnotesize]{}  (X);
\draw[->, line width= 1] (Ng) -- node[above,font=\footnotesize]{}  (X);
\draw[->, line width= 1] (V) -- node[above,font=\footnotesize]{}  (Gg);
\draw[->, line width= 1] (V) -- node[above,font=\footnotesize]{}  (Ng);
\draw[->, line width= 1] (Nb) -- node[above,font=\footnotesize]{}  (B);
\draw[->, line width= 1] (Gb) -- node[above,font=\footnotesize]{}  (B);
\draw[->, line width= 1] (B) -- node[above,font=\footnotesize]{}  (X);
\draw[->, line width= 1] (B) -- node[above,font=\footnotesize]{\quad \quad \quad $\eta_b, \gamma_b$}  (Y);
\draw[->, line width= 1] (V) -- node[above,font=\footnotesize]{}  (B);
\draw[->, line width= 1] (Ylag) -- node[above,font=\footnotesize]{}  (B);
\draw[->, line width=1] (V) to  [out=270,in=270, looseness=1.5] node[below, font=\footnotesize]{$\nu$} (Y);
\draw[->, line width=1] (Ylag) to  [out=0,in=90, looseness=1.5] node[below, font=\footnotesize]{$\boldsymbol{\phi}$} (Y);

\end{tikzpicture}
\caption*{\footnotesize The complete version of the DAG suggests how mobility ($X$) interacts with vaccination ($V$), Google Trends ($G_b$), News ($N_b$) and behavioral effects ($B$). Each arrow suggests a connection between two variables. Following \cite{elwert2013graphical}, only missing arrows makes assumptions regarding causal relations. The circle around the behavioral variable ($B$) denotes a non-observable variable, while $Y_{lag}$ denotes a 4-dimensional vector of lagged observations of Covid-19 cases (deaths). Also, Google Trends general searches ($G_g$) and general News ($N_b$) are related to vaccination ($V$) and mobility ($X$).}
\end{figure}

\begin{figure}[H]

    \caption{DAG for 2020: shutting down vaccination channel}
    \label{fig:dag2020}
    \centering
    \begin{tikzpicture}
\node[text centered] (X) {$X$};
\node[above left = 1.5 of X, text centered] (Ng) {$N_{g}$};
\node[below=1 of Ng, text centered] (Gg) {$G_{g}$};
\node[above=1 of X, text centered] (B) {$\mycirc{$B$}$};
\node[above left=0.5 of B, text centered] (Gb) {$G_{b}$};
\node[above=1 of B, text centered] (Ylag) {$Y_{\text{lag}}$};
\node[above right = 0.5 of B, text centered] (Nb) {$N_{b}$};
\node[right = 3 of X, text centered] (Y) {$Y$};

\draw[->, line width= 1] (X) -- node[above,font=\footnotesize]{$\boldsymbol{\beta}$}  (Y);
\draw[->, line width= 1] (Gg) -- node[above,font=\footnotesize]{}  (X);
\draw[->, line width= 1] (Ng) -- node[above,font=\footnotesize]{}  (X);
\draw[->, line width= 1] (Nb) -- node[above,font=\footnotesize]{}  (B);
\draw[->, line width= 1] (Gb) -- node[above,font=\footnotesize]{}  (B);
\draw[->, line width= 1] (B) -- node[above,font=\footnotesize]{}  (X);
\draw[->, line width= 1] (B) -- node[above,font=\footnotesize]{\quad \quad \quad $\eta_b, \gamma_b$}  (Y);
\draw[->, line width= 1] (Ylag) -- node[above,font=\footnotesize]{}  (B);
\draw[->, line width=1] (Ylag) to  [out=0,in=90, looseness=1.5] node[below, font=\footnotesize]{$\boldsymbol{\phi}$} (Y);

\end{tikzpicture}
\caption*{\footnotesize The shrinked version of the DAG (only for 2020) shuts down the vaccination channel that affects mobility, behavior and cases (deaths). Such transformation would imply in a cleaner identification of mobility effect, but with local validation only, not extending to the subsequent second wave of Covid-19 spread.}
\end{figure}

\bigskip

\bigskip
\begin{figure}[H]
\caption{Median Number of Days from First Symptom to Obit}
    \label{fig:death}
    \centering
    \includegraphics[width=16cm, height = 5cm]{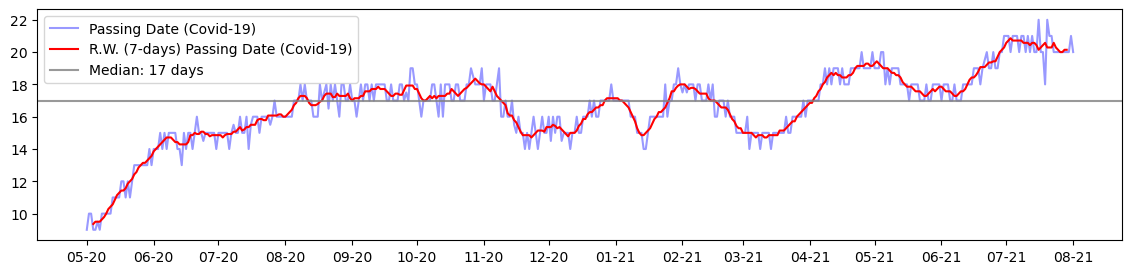}
    \caption*{\footnotesize The plot reveals the median number of days taken from first symptom up to obit from patients that are considered on the SRAG data over the period of May, 2020 up to August, 2021. We also represent a 7-days rolling window (R.W.) as a smoothing and the overall median (17 days). This time series is important while defining the number of lags to consider on the estimation.}
    \label{fig:death}
\end{figure}


\clearpage

\section*{Appendix A. Introduction to Direct Acyclic Graphs} \label{sec:appendixa}
\addcontentsline{toc}{section}{Appendix A}

The usage of DAGs in causal inference have roots in \cite{pearl1995causal} and \cite{pearl2009causality}, providing causal interpretation based on variables relations. Based on our estimation, the principal objective is to generate identification of the causal effect of mobility on Covid-19 cases and deaths. Our aim is to isolate causal from noncausal associations. As described by \cite{elwert2013graphical}, DAGs are a powerful tool to identify what control variables should we include and which we should not include to ``achieve identification".

The main idea of using DAGs consists in generating a clear and objective graph that should encode the main causal relations that we aim to describe in our model. In a glance, we should interpret a DAG based on three elements, following \cite{elwert2013graphical}: (i) Variables, that are represented in nodes; (ii) Arrows, suggesting possible direct causal impacts; (iii) Missing arrows, encoding ``sharp assumptions" about absence of causality effects.

Ideally, if we were able to observe mobility in a way that it is not affected by external elements, i.e. exogenously (e.g. a randomized version of mobility), we would retrieve the causal effect of mobility on Covid-19 variables in a direct way. This representation would imply in the Figure \ref{fig:dag1} Graph:

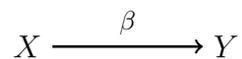
\begin{figure}[H]
\caption{DAG representing causal chain between mobility and cases (deaths)}
    \label{fig:dag1}
    \centering
      \begin{tikzpicture}
\node[text centered] (X) {$X$};
\node[right = 2 of X, text centered] (Y) {$Y$};

\draw[->, line width= 1] (X) -- node[above,font=\footnotesize]{$\beta$}  (Y);

\end{tikzpicture}
\end{figure}

As we do not observe this ``artificial" measure of mobility, we cannot retrieve a causal effect merely by regressing solely those two variables. Therefore, we should include potential control variables that are related to mobility measures in order to remove the omission bias. As an example, take vaccination as a potential control: this variable affect both mobility and number of cases and deaths. Therefore, its inclusion as an regressor (estimating $\eta$) is necessary to correctly identify the causal effect of mobility, i.e. without vaccination we would bias the estimation of $\beta$ through the channel of the $\eta$ relation. This result on the Figure \ref{fig:dag2} DAG:

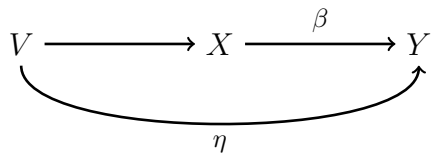
\begin{figure}[H]
\caption{DAG representing causal chain between vaccination, mobility and cases (deaths)}
    \label{fig:dag2}
    \centering
  \begin{tikzpicture}
\node[text centered] (X) {$X$};
\node[right = 2 of X, text centered] (Y) {$Y$};
\node[left = 2 of X, text centered] (V) {$V$};

\draw[->, line width= 1] (V) -- node[above,font=\footnotesize]{}  (X);
\draw[->, line width= 1] (X) -- node[above,font=\footnotesize]{$\beta$}  (Y);
\draw[->, line width=1] (V) to  [out=270,in=270, looseness=0.5] node[below, font=\footnotesize]{$\eta$} (Y);

\end{tikzpicture}
\end{figure}

By proceeding in the same fashion, we identify potential variables that are useful controls in our identification strategy, as Google Trends searches and news related to Covid-19. The result is the complete DAG that we present on Section \ref{sec:ident}.

\clearpage

\section*{Appendix B. Keywords and Categories} \label{sec:appendixb}
\addcontentsline{toc}{section}{Appendix B}

To generate our soft data controls, we need to specify the Google Trends (gt-series) search terms chosen by individuals or the keywords used to constitute the News-Index (n-index). The objective is to be concise and precise while selecting the terms in order to capture a general overview of the Covid-19 situation at the individual level that may affect both mobility and number of cases (deaths), according to the DAG proposed in Figures \ref{fig:dag2021} and \ref{fig:dag2020}. We also focus in generating compatibility within search terms and keywords for both controls.

There are two main categories for both gt and n-series that encode general effects (g) and behavioral effects (b). To capture general effects (g) of the Covid-19 pandemy, we selected words that refer to: (i) Covid-19 related terms; (ii) Fake news terms; (iii) Vaccination terms. The first topic has been chosen with the objective of getting the overall trend of the number of infections over time. The second topic has the objective of allowing the direct impact of fake news spread over Covid-19 evolution. The third topic is related to the vaccination campaign to correctly control the infection evolution.
\begin{enumerate}
    \item General Covid-19 related terms: covid, pandemia, coronavirus, covid-19, mortes covid, morrer de covid, covid o que fazer, covid como proceder, pegar covid, transmissão covid, covid mata, covid contagioso, covid transmite, contagio covid, sintomas covid, morte de covid, casos covid.
    \item Fake News related terms: kit-covid, hidroxicloroquina, cloroquina, azitromicina, gripezinha, ivermectina, remedio covid, tratamento covid.
    \item Vaccination related terms: vacinação covid, vacinas covid, pfizer, astrazeneca, janssen, butantan, coronavac, moderna, biontech, oxford, fiocruz, sputnik v.
\end{enumerate}

To capture the effects of the behavioral category (b), we selected prevention related terms. This category should embed individual behavior (not observed by hard data indicators) that may affect the infection evolution throughout time. As behavior is a non-observable variable, the absence of such terms would bias the effects of mobility on cases (deaths).

\begin{enumerate}
\setcounter{enumi}{3}
    \item Prevention related terms: mascara, lavas as mãos, alcool em gel, isolamento, distanciamento, quarentena, lockdown, confinamento, ficar em casa, toque de recolher, toque de restrição, restrições, circulação.
\end{enumerate}

In terms of collecting Google searches, the Google Trends API for R already supplies the number of counts for a specified period for each search term given. To create the News-Index, we recur to a similar approach as in \cite{baker2016measuring}, collecting G1 news regarding Covid-19 and counting the topic appearances based on the presence or absence of the term in the title of the news. This way, we solely focus on news that focus on the given topic directly\footnote{Differently, we could have count the number of direct and indirect appearances of a given topic based on its inclusion on overall text elements. However, by focusing on its appearance on the title, we restrict the number of news to consider only directly refer topics.}.

Formally, as the sets of search words and keywords used to create gt and n-series are equal, we will define the indexes suppressing the gt or n marker. Recall that we have a panel data (at state level and weekly frequency) of each topic. Therefore, each observation is indexed by state marker $j \in \{1,\cdots,27\}$ and a time stamp $t \in \{t_0, \cdots, T\}$.

For each gt and n-series of controls, we have a vector $\boldsymbol{g_{(j,t)}} = (g_{1,(j,t)},g_{2,(j,t)},g_{3,(j,t)},g_{4,(j,t)})$ that represents the four above categories. For each category $g_{i,(j,t)}$, for $i \in \{1,2,3,4\}$, there are $n_{i,(j,t)}$ associated search words (for gt) or keywords (for n), denoted $g_{i,w,(j,t)}$. We thus generate the indexes in the following manner:

\begin{equation}
    g_{i,(j,t)} = \displaystyle \sum_{w = 1}^{n_{i,(j,t)}} g_{i,w,(j,t)}
\end{equation}

\noindent for each $i \in \{1,2,3,4\}$, $j \in \{1,\cdots,27\}$ and $t \in \{t_0, \cdots, T\}$.

As an example, consider the News-Index (n) regarding prevention related terms, i.e. $i = 4$. In this case, we have $n_{4,(j,t)} = 13$ and $w$ denotes a certain topic (e.g. $w = \text{mascara}$). Therefore, the $g_{4,\text{mascara},(j,t)}$ denotes the number of counts of ``mascara'', that belongs to prevention related terms, over each state $j$ and each week $t$. Finally, the News-Index for prevention related terms is given by:

\begin{equation*}
    g_{4,(j,t)} = \displaystyle \sum_{w = 1}^{13} g_{4,w,(j,t)}
\end{equation*}

\clearpage

\section*{Appendix C. First Stage Estimations} \label{sec:appendixc}
\addcontentsline{toc}{section}{Appendix C}

\begin{table}[H]
\centering
\caption{First Stage Estimation for Mobility Measures against gt-series, n-index and lagged Covid-19 variables\label{tab:1stage}}
\begin{adjustbox}{max width= 16cm}
\begin{threeparttable}
{
\def\sym#1{\ifmmode^{#1}\else\(^{#1}\)\fi}
\begin{tabular}{l*{12}{c}}
\hline\hline
            &\multicolumn{1}{c}{(1)}&\multicolumn{1}{c}{(2)}&\multicolumn{1}{c}{(3)}&\multicolumn{1}{c}{(4)}&\multicolumn{1}{c}{(5)}&\multicolumn{1}{c}{(6)}&\multicolumn{1}{c}{(7)}&\multicolumn{1}{c}{(8)}&\multicolumn{1}{c}{(9)}&\multicolumn{1}{c}{(10)}&\multicolumn{1}{c}{(11)}&\multicolumn{1}{c}{(12)}\\
            &\multicolumn{1}{c}{residential}&\multicolumn{1}{c}{residential}&\multicolumn{1}{c}{residential}&\multicolumn{1}{c}{residential}&\multicolumn{1}{c}{residential}&\multicolumn{1}{c}{residential}&\multicolumn{1}{c}{workplace}&\multicolumn{1}{c}{workplace}&\multicolumn{1}{c}{workplace}&\multicolumn{1}{c}{workplace}&\multicolumn{1}{c}{workplace}&\multicolumn{1}{c}{workplace}\\
\hline
n\_covid     &     0.00351\sym{***}&                     &                     &                     &                     &                     &     0.00307\sym{***}&                     &                     &                     &                     &                     \\
            &  (0.000180)         &                     &                     &                     &                     &                     &  (0.000504)         &                     &                     &                     &                     &                     \\
[1em]
n\_fakenews  &      0.0256\sym{*}  &                     &                     &                     &                     &                     &      -0.172\sym{***}&                     &                     &                     &                     &                     \\
            &    (0.0104)         &                     &                     &                     &                     &                     &    (0.0286)         &                     &                     &                     &                     &                     \\
[1em]
n\_vaccines  &     0.00338\sym{***}&                     &                     &                     &                     &                     &    -0.00865\sym{***}&                     &                     &                     &                     &                     \\
            &  (0.000258)         &                     &                     &                     &                     &                     &  (0.000733)         &                     &                     &                     &                     &                     \\
[1em]
n\_prevention&                     &      0.0207\sym{***}&                     &                     &                     &                     &                     &    -0.00934\sym{***}&                     &                     &                     &                     \\
            &                     &  (0.000765)         &                     &                     &                     &                     &                     &   (0.00216)         &                     &                     &                     &                     \\
[1em]
gt\_covid    &                     &                     &     0.00136\sym{***}&                     &                     &                     &                     &                     &    -0.00475\sym{***}&                     &                     &                     \\
            &                     &                     & (0.0000633)         &                     &                     &                     &                     &                     &  (0.000170)         &                     &                     &                     \\
[1em]
gt\_fakenews &                     &                     &     0.00456\sym{***}&                     &                     &                     &                     &                     &     -0.0125\sym{***}&                     &                     &                     \\
            &                     &                     &  (0.000227)         &                     &                     &                     &                     &                     &  (0.000589)         &                     &                     &                     \\
[1em]
gt\_vaccines &                     &                     &   -0.000341\sym{*}  &                     &                     &                     &                     &                     &     0.00378\sym{***}&                     &                     &                     \\
            &                     &                     &  (0.000152)         &                     &                     &                     &                     &                     &  (0.000409)         &                     &                     &                     \\
[1em]
gt\_prevention&                     &                     &                     &     0.00241\sym{***}&                     &                     &                     &                     &                     &    -0.00261\sym{***}&                     &                     \\
            &                     &                     &                     &  (0.000136)         &                     &                     &                     &                     &                     &  (0.000362)         &                     &                     \\
[1em]
cases\_1     &                     &                     &                     &                     &     -0.0819\sym{***}&                     &                     &                     &                     &                     &      0.0712         &                     \\
            &                     &                     &                     &                     &    (0.0139)         &                     &                     &                     &                     &                     &    (0.0411)         &                     \\
[1em]
cases\_2     &                     &                     &                     &                     &      0.0721\sym{***}&                     &                     &                     &                     &                     &      -0.286\sym{***}&                     \\
            &                     &                     &                     &                     &    (0.0152)         &                     &                     &                     &                     &                     &    (0.0454)         &                     \\
[1em]
cases\_3     &                     &                     &                     &                     &       0.247\sym{***}&                     &                     &                     &                     &                     &      -0.625\sym{***}&                     \\
            &                     &                     &                     &                     &    (0.0151)         &                     &                     &                     &                     &                     &    (0.0450)         &                     \\
[1em]
cases\_4     &                     &                     &                     &                     &       0.261\sym{***}&                     &                     &                     &                     &                     &      -0.611\sym{***}&                     \\
            &                     &                     &                     &                     &    (0.0137)         &                     &                     &                     &                     &                     &    (0.0406)         &                     \\
[1em]
deaths\_1    &                     &                     &                     &                     &                     &       0.344\sym{***}&                     &                     &                     &                     &                     &      -0.758\sym{***}\\
            &                     &                     &                     &                     &                     &    (0.0156)         &                     &                     &                     &                     &                     &    (0.0518)         \\
[1em]
deaths\_2    &                     &                     &                     &                     &                     &       0.547\sym{***}&                     &                     &                     &                     &                     &      -1.213\sym{***}\\
            &                     &                     &                     &                     &                     &    (0.0177)         &                     &                     &                     &                     &                     &    (0.0588)         \\
[1em]
deaths\_3    &                     &                     &                     &                     &                     &       0.545\sym{***}&                     &                     &                     &                     &                     &      -1.250\sym{***}\\
            &                     &                     &                     &                     &                     &    (0.0176)         &                     &                     &                     &                     &                     &    (0.0582)         \\
[1em]
deaths\_4    &                     &                     &                     &                     &                     &       0.362\sym{***}&                     &                     &                     &                     &                     &      -0.818\sym{***}\\
            &                     &                     &                     &                     &                     &    (0.0154)         &                     &                     &                     &                     &                     &    (0.0509)         \\
[1em]
\_cons      &      -0.588\sym{***}&      -0.572\sym{***}&       3.177\sym{***}&       3.182\sym{***}&      -1.554\sym{***}&      -5.141\sym{***}&       15.13\sym{***}&       15.19\sym{***}&      -17.08\sym{***}&      -17.48\sym{***}&       16.37\sym{***}&       25.83\sym{***}\\
            &    (0.0560)         &    (0.0560)         &    (0.0572)         &    (0.0589)         &     (0.326)         &     (1.471)         &     (0.174)         &     (0.174)         &     (0.161)         &     (0.164)         &     (1.193)         &     (5.277)         \\
\hline
$R^2$          &       0.802         &       0.801         &       0.808         &       0.801         &       0.730         &       0.727         &       0.727         &       0.727         &       0.726         &       0.719         &       0.673         &       0.722         \\
$R_{adj}^2$        &       0.798         &       0.798         &       0.805         &       0.798         &       0.723         &       0.716         &       0.723         &       0.723         &       0.722         &       0.714         &       0.661         &       0.705         \\
N           &       69734         &       69734         &       70641         &       70641         &       45107         &       26576         &      156881         &      156881         &      158809         &      158809         &       64493         &       30896         \\
p           &           0         &           0         &           0         &           0         &           0         &           0         &           0         &           0         &           0         &           0         &           0         &           0         \\
\hline\hline
\multicolumn{13}{l}{\footnotesize Standard errors in parentheses}\\
\multicolumn{13}{l}{\footnotesize \sym{*} \(p<0.05\), \sym{**} \(p<0.01\), \sym{***} \(p<0.001\)}\\
\end{tabular}
}

\begin{tablenotes}
\small
\item[a]{Results for Fixed-Effects Estimation over 2020 and 2021 sample, aggregating cases (deaths) by notification area. The first stage estimation has the objective of validating the causal relations suggested on the DAG present on Figure \ref{fig:dag2021}. $R^2$ denotes the R-squared, $R_{adj}^2$ the adjusted R-squared, $N$ the total number of observations used on the estimation and $p$ the F-test associated p-value.}
\end{tablenotes}
\end{threeparttable}
\end{adjustbox}
\end{table}

\begin{table}[H]
\centering
\caption{First Stage Estimation for Cases (Deaths) against Vaccination Campaign\label{tab:vac/cases}}
\begin{adjustbox}{max width= 16cm}
\begin{threeparttable}
\small
{
\def\sym#1{\ifmmode^{#1}\else\(^{#1}\)\fi}
\begin{tabular}{l*{14}{c}}
\hline\hline
            &\multicolumn{1}{c}{(1)}&\multicolumn{1}{c}{(2)}&\multicolumn{1}{c}{(3)}&\multicolumn{1}{c}{(4)}&\multicolumn{1}{c}{(5)}&\multicolumn{1}{c}{(6)}&\multicolumn{1}{c}{(7)}&\multicolumn{1}{c}{(8)}&\multicolumn{1}{c}{(9)}&\multicolumn{1}{c}{(10)}&\multicolumn{1}{c}{(11)}&\multicolumn{1}{c}{(12)}&\multicolumn{1}{c}{(13)}&\multicolumn{1}{c}{(14)}\\
            &\multicolumn{1}{c}{cases}&\multicolumn{1}{c}{cases}&\multicolumn{1}{c}{cases}&\multicolumn{1}{c}{cases}&\multicolumn{1}{c}{cases}&\multicolumn{1}{c}{cases}&\multicolumn{1}{c}{cases}&\multicolumn{1}{c}{cases}&\multicolumn{1}{c}{deaths}&\multicolumn{1}{c}{deaths}&\multicolumn{1}{c}{deaths}&\multicolumn{1}{c}{deaths}&\multicolumn{1}{c}{deaths}&\multicolumn{1}{c}{deaths}\\
            &         m=1         &         m=1         &         m=2         &         m=2         &         m=3         &         m=3         &         m=4         &         m=4         &         m=2         &         m=2         &         m=3         &         m=3         &         m=4         &         m=4         \\
\hline
1st\_dose    &     0.00147         &                     &    -0.00832\sym{*}  &                     &    0.000469         &                     &    -0.00476         &                     &   -0.000159         &                     &     0.00904         &                     &     -0.0116\sym{*}  &                     \\
            &   (0.00334)         &                     &   (0.00345)         &                     &   (0.00353)         &                     &   (0.00375)         &                     &   (0.00476)         &                     &   (0.00480)         &                     &   (0.00507)         &                     \\
[1em]
2nd\_dose    &     0.00237         &                     &    -0.00541\sym{*}  &                     &   -0.000967         &                     &    -0.00452         &                     &    -0.00200         &                     &    -0.00808\sym{*}  &                     &     0.00351         &                     \\
            &   (0.00238)         &                     &   (0.00244)         &                     &   (0.00243)         &                     &   (0.00252)         &                     &   (0.00339)         &                     &   (0.00333)         &                     &   (0.00348)         &                     \\
[1em]
srag\_vac    &                     &      -0.226\sym{***}&                     &     -0.0511\sym{***}&                     &     -0.0497\sym{***}&                     &     -0.0317\sym{***}&                     &      0.0147         &                     &     -0.0198\sym{*}  &                     &     -0.0569\sym{***}\\
            &                     &   (0.00818)         &                     &   (0.00710)         &                     &   (0.00760)         &                     &   (0.00758)         &                     &   (0.00887)         &                     &   (0.00907)         &                     &   (0.00979)         \\
[1em]
\_cons      &       0.663\sym{***}&      -0.331\sym{***}&       0.256\sym{***}&     -0.0521         &       0.428\sym{***}&      -0.106         &      0.0326         &       0.123         &       1.064\sym{***}&      0.0898         &       0.215\sym{***}&       0.111         &       0.273\sym{***}&      -0.167         \\
            &    (0.0210)         &    (0.0774)         &    (0.0221)         &    (0.0540)         &    (0.0232)         &    (0.0696)         &    (0.0230)         &    (0.0654)         &    (0.0264)         &    (0.0738)         &    (0.0274)         &    (0.0773)         &    (0.0290)         &    (0.0888)         \\
\hline
$R^2$          &      0.0304         &      0.0771         &      0.0296         &      0.0372         &      0.0283         &      0.0320         &      0.0247         &      0.0367         &      0.0237         &      0.0308         &      0.0229         &      0.0228         &      0.0216         &      0.0225         \\
$R_{overall}^2$        &      0.0277         &      0.0126         &      0.0274         &      0.0269         &      0.0260         &      0.0243         &      0.0235         &      0.0287         &      0.0204         &      0.0240         &      0.0195         &      0.0194         &      0.0185         &      0.0169         \\
N           &       57802         &       25082         &       55725         &       21893         &       53420         &       20379         &       50964         &       18951         &       32101         &       16367         &       31124         &       15455         &       29951         &       14213         \\
p           &           .         &   3.45e-279         &           .         &   6.23e-198         &           .         &   2.93e-153         &           .         &   7.29e-159         &           .         &   7.71e-145         &           .         &    1.25e-92         &           .         &    2.83e-88         \\
\hline\hline
\multicolumn{15}{l}{\footnotesize Standard errors in parentheses}\\
\multicolumn{15}{l}{\footnotesize \sym{*} \(p<0.05\), \sym{**} \(p<0.01\), \sym{***} \(p<0.001\)}\\
\end{tabular}
}

\begin{tablenotes}
\small
\item[a]{Results for Fixed-Effects Estimation over 2020 and 2021 sample. We regress cases of Covid-19 taken by residence area against Covid-19 vaccination evolution also taken by residence area. The first stage estimation has the objective of validating the causal relations suggested on the DAG present on Figure \ref{fig:dag2021}. $R^2$ denotes the R-squared, $R_{overall}^2$ the overall R-squared, $N$ the total number of observations used on the estimation and $p$ the F-test associated p-value.}
\end{tablenotes}
\end{threeparttable}
\end{adjustbox}
\end{table}

\begin{table}[H]
\small
\centering
\caption{First Stage Estimation for Mobility against Vaccination Campaign\label{tab:vac/mob}}
\begin{adjustbox}{max width= 16cm}
\begin{threeparttable}
\small
{
\def\sym#1{\ifmmode^{#1}\else\(^{#1}\)\fi}
\begin{tabular}{l*{14}{c}}
\hline\hline
            &\multicolumn{1}{c}{(1)}&\multicolumn{1}{c}{(2)}&\multicolumn{1}{c}{(3)}&\multicolumn{1}{c}{(4)}&\multicolumn{1}{c}{(5)}&\multicolumn{1}{c}{(6)}&\multicolumn{1}{c}{(7)}&\multicolumn{1}{c}{(8)}&\multicolumn{1}{c}{(9)}&\multicolumn{1}{c}{(10)}&\multicolumn{1}{c}{(11)}&\multicolumn{1}{c}{(12)}&\multicolumn{1}{c}{(13)}&\multicolumn{1}{c}{(14)}\\
            &\multicolumn{1}{c}{residential}&\multicolumn{1}{c}{residential}&\multicolumn{1}{c}{residential}&\multicolumn{1}{c}{residential}&\multicolumn{1}{c}{residential}&\multicolumn{1}{c}{residential}&\multicolumn{1}{c}{residential}&\multicolumn{1}{c}{residential}&\multicolumn{1}{c}{workplace}&\multicolumn{1}{c}{workplace}&\multicolumn{1}{c}{workplace}&\multicolumn{1}{c}{workplace}&\multicolumn{1}{c}{workplace}&\multicolumn{1}{c}{workplace}\\
            &         m=1         &         m=1         &         m=2         &         m=2         &         m=3         &         m=3         &         m=4         &         m=4         &         m=2         &         m=2         &         m=3         &         m=3         &         m=4         &         m=4         \\
\hline
1st\_dose    &    -0.00220         &                     &    -0.00815         &                     &    -0.00963         &                     &     -0.0189         &                     &       0.151\sym{**} &                     &       0.135\sym{**} &                     &       0.132\sym{**} &                     \\
            &    (0.0173)         &                     &    (0.0176)         &                     &    (0.0175)         &                     &    (0.0172)         &                     &    (0.0478)         &                     &    (0.0466)         &                     &    (0.0464)         &                     \\
[1em]
2nd\_dose    &      0.0189         &                     &      0.0165         &                     &      0.0185         &                     &      0.0165         &                     &      0.0688\sym{*}  &                     &      0.0546\sym{*}  &                     &      0.0471         &                     \\
            &   (0.00974)         &                     &   (0.00996)         &                     &    (0.0102)         &                     &    (0.0104)         &                     &    (0.0270)         &                     &    (0.0271)         &                     &    (0.0270)         &                     \\
[1em]
srag\_vac    &                     &       0.230\sym{***}&                     &       0.238\sym{***}&                     &       0.234\sym{***}&                     &       0.256\sym{***}&                     &      -0.567\sym{***}&                     &      -0.514\sym{***}&                     &      -0.578\sym{***}\\
            &                     &    (0.0321)         &                     &    (0.0332)         &                     &    (0.0345)         &                     &    (0.0355)         &                     &     (0.101)         &                     &     (0.102)         &                     &     (0.104)         \\
[1em]
\_cons      &       10.93\sym{***}&       6.406\sym{***}&       10.85\sym{***}&       6.386\sym{***}&       10.82\sym{***}&       6.373\sym{***}&       10.71\sym{***}&       6.371\sym{***}&      -15.17\sym{***}&      -3.537\sym{***}&      -15.21\sym{***}&      -3.511\sym{***}&      -15.15\sym{***}&      -3.496\sym{***}\\
            &    (0.0780)         &     (0.304)         &    (0.0795)         &     (0.309)         &    (0.0786)         &     (0.315)         &    (0.0773)         &     (0.318)         &     (0.270)         &     (0.927)         &     (0.262)         &     (0.913)         &     (0.259)         &     (0.896)         \\
\hline
$R^2$          &       0.522         &       0.567         &       0.513         &       0.550         &       0.508         &       0.539         &       0.512         &       0.542         &       0.585         &       0.648         &       0.579         &       0.655         &       0.576         &       0.657         \\
$R_{overall}^2$        &       0.165         &       0.208         &       0.158         &       0.199         &       0.156         &       0.190         &       0.155         &       0.191         &       0.350         &       0.403         &       0.344         &       0.408         &       0.340         &       0.413         \\
N           &       24398         &       13225         &       23387         &       12571         &       22373         &       11896         &       21363         &       11203         &       48193         &       18864         &       46066         &       17778         &       43941         &       16650         \\
p           &           .         &           0         &           .         &           0         &           .         &           0         &           .         &           0         &           .         &           0         &           .         &           0         &           .         &           0         \\
\hline\hline
\multicolumn{15}{l}{\footnotesize Standard errors in parentheses}\\
\multicolumn{15}{l}{\footnotesize \sym{*} \(p<0.05\), \sym{**} \(p<0.01\), \sym{***} \(p<0.001\)}\\
\end{tabular}
}

\begin{tablenotes}
\small
\item[a]{Results for Fixed-Effects Estimation over 2020 and 2021 sample. We regress deaths of Covid-19 taken by residence area against Covid-19 vaccination evolution also taken by residence area. The first stage estimation has the objective of validating the causal relations suggested on the DAG present on Figure \ref{fig:dag2021}.$R^2$ denotes the R-squared, $R_{overall}^2$ the overall R-squared, $N$ the total number of observations used on the estimation and $p$ the F-test associated p-value.}
\end{tablenotes}
\end{threeparttable}
\end{adjustbox}
\end{table}

\section*{Appendix D. Robustness Estimations} \label{sec:appendixd}
\addcontentsline{toc}{section}{Appendix D}

\begin{table}[H]
\centering
\caption{Estimation for Complete Sample (2020-2021) for all Mobility Measures\label{tab:moball}}
\begin{adjustbox}{max width= 10cm}
\begin{threeparttable}
{
\def\sym#1{\ifmmode^{#1}\else\(^{#1}\)\fi}
\begin{tabular}{l*{7}{c}}
\hline\hline
            &\multicolumn{1}{c}{(1)}&\multicolumn{1}{c}{(2)}&\multicolumn{1}{c}{(3)}&\multicolumn{1}{c}{(4)}&\multicolumn{1}{c}{(5)}&\multicolumn{1}{c}{(6)}&\multicolumn{1}{c}{(7)}\\
            &\multicolumn{1}{c}{cases}&\multicolumn{1}{c}{cases}&\multicolumn{1}{c}{cases}&\multicolumn{1}{c}{cases}&\multicolumn{1}{c}{deaths}&\multicolumn{1}{c}{deaths}&\multicolumn{1}{c}{deaths}\\
            &         m=1         &         m=2         &         m=3         &         m=4         &         m=2         &         m=3         &         m=4         \\
\hline
residential &     -0.0656\sym{***}&     -0.0499\sym{***}&     -0.0461\sym{***}&     -0.0233\sym{***}&     -0.0219\sym{*}  &     -0.0538\sym{***}&     -0.0629\sym{***}\\
            &   (0.00632)         &   (0.00631)         &   (0.00631)         &   (0.00676)         &   (0.00876)         &   (0.00817)         &   (0.00868)         \\
[1em]
workplace   &    -0.00330         &    -0.00198         &    -0.00209         &     0.00313         &     0.00200         &    -0.00136         &    0.000866         \\
            &   (0.00186)         &   (0.00194)         &   (0.00192)         &   (0.00204)         &   (0.00226)         &   (0.00235)         &   (0.00265)         \\
[1em]
parks       &   -0.000301         &   -0.000750         &   -0.000344         &    0.000475         &    0.000959         &    0.000266         &   -0.000353         \\
            &  (0.000471)         &  (0.000469)         &  (0.000466)         &  (0.000505)         &  (0.000631)         &  (0.000686)         &  (0.000716)         \\
[1em]
grocery     &    -0.00286\sym{**} &    -0.00287\sym{**} &    -0.00438\sym{***}&    -0.00316\sym{***}&    -0.00242         &    -0.00265\sym{*}  &    -0.00321\sym{*}  \\
            &   (0.00100)         &  (0.000951)         &   (0.00101)         &  (0.000909)         &   (0.00128)         &   (0.00130)         &   (0.00133)         \\
[1em]
transit     &    -0.00133\sym{*}  &    -0.00102         &    -0.00130\sym{*}  &  -0.0000425         &    -0.00160         &    -0.00203\sym{*}  &    -0.00265\sym{**} \\
            &  (0.000637)         &  (0.000600)         &  (0.000632)         &  (0.000686)         &  (0.000821)         &  (0.000913)         &  (0.000971)         \\
[1em]
retail      &     0.00425\sym{**} &     0.00407\sym{***}&     0.00282\sym{*}  &    -0.00241         &     0.00461\sym{*}  &     0.00571\sym{**} &     0.00532\sym{**} \\
            &   (0.00137)         &   (0.00115)         &   (0.00124)         &   (0.00124)         &   (0.00181)         &   (0.00180)         &   (0.00183)         \\
[1em]
1st\_dose    &     0.00564         &   -0.000776         &    -0.00648         &    0.000654         &     0.00485         &     0.00922         &     0.00723         \\
            &   (0.00581)         &   (0.00643)         &   (0.00632)         &   (0.00673)         &   (0.00872)         &   (0.00869)         &   (0.00904)         \\
[1em]
2nd\_dose    &     0.00924\sym{*}  &    -0.00734\sym{*}  &    0.000751         &    -0.00326         &    -0.00810         &    -0.00765         &    -0.00101         \\
            &   (0.00415)         &   (0.00355)         &   (0.00389)         &   (0.00440)         &   (0.00582)         &   (0.00591)         &   (0.00552)         \\
[1em]
n\_covid     &    0.000946\sym{***}&    0.000568\sym{**} &    0.000371         &    0.000737\sym{***}&    0.000747\sym{*}  &    0.000428         &    0.000394         \\
            &  (0.000217)         &  (0.000212)         &  (0.000229)         &  (0.000211)         &  (0.000319)         &  (0.000309)         &  (0.000300)         \\
[1em]
n\_prevention&  -0.0000170         &    0.000559         &   -0.000322         &    -0.00105         &    -0.00369\sym{**} &     0.00122         &     0.00198         \\
            &  (0.000948)         &  (0.000984)         &  (0.000868)         &   (0.00102)         &   (0.00124)         &   (0.00131)         &   (0.00129)         \\
[1em]
n\_fakenews  &     -0.0174\sym{**} &     -0.0173\sym{**} &    0.000936         &    0.000556         &     -0.0165         &     -0.0187\sym{*}  &     -0.0224\sym{**} \\
            &   (0.00532)         &   (0.00581)         &   (0.00623)         &   (0.00613)         &   (0.00873)         &   (0.00826)         &   (0.00800)         \\
[1em]
n\_vaccines  &   -0.000689\sym{*}  &    -0.00142\sym{***}&    -0.00154\sym{***}&   -0.000897\sym{*}  &   -0.000132         &   -0.000561         &   -0.000863         \\
            &  (0.000316)         &  (0.000333)         &  (0.000374)         &  (0.000374)         &  (0.000428)         &  (0.000479)         &  (0.000514)         \\
[1em]
gt\_covid    &  -0.0000288         &   -0.000260\sym{***}&   -0.000393\sym{***}&   -0.000502\sym{***}&    0.000484\sym{***}&    0.000132         &   0.0000709         \\
            & (0.0000535)         & (0.0000543)         & (0.0000529)         & (0.0000609)         & (0.0000861)         & (0.0000782)         & (0.0000739)         \\
[1em]
gt\_prevention&    0.000139         &    0.000246\sym{*}  &   0.0000858         & -0.00000878         &   0.0000138         &   0.0000244         &  -0.0000168         \\
            &  (0.000119)         &  (0.000118)         &  (0.000116)         &  (0.000137)         &  (0.000175)         &  (0.000161)         &  (0.000164)         \\
[1em]
gt\_fakenews &  -0.0000890         &   -0.000237         &   -0.000158         &    0.000140         &    0.000245         &    0.000264         &   -0.000554\sym{*}  \\
            &  (0.000164)         &  (0.000164)         &  (0.000166)         &  (0.000157)         &  (0.000235)         &  (0.000226)         &  (0.000217)         \\
[1em]
gt\_vaccines &  -0.0000675         &    0.000187         &    0.000245\sym{*}  &    0.000326\sym{**} &   -0.000529\sym{***}&   -0.000223         &   -0.000463\sym{**} \\
            &  (0.000103)         &  (0.000107)         &  (0.000111)         &  (0.000118)         &  (0.000140)         &  (0.000154)         &  (0.000160)         \\
[1em]
cases\_1  &      -0.435\sym{***}&      -0.436\sym{***}&      -0.450\sym{***}&      -0.448\sym{***}&                     &                     &                     \\
            &    (0.0154)         &    (0.0157)         &    (0.0162)         &    (0.0172)         &                     &                     &                     \\
[1em]
cases\_2  &      -0.176\sym{***}&      -0.198\sym{***}&      -0.221\sym{***}&      -0.212\sym{***}&                     &                     &                     \\
            &    (0.0157)         &    (0.0168)         &    (0.0169)         &    (0.0183)         &                     &                     &                     \\
[1em]
cases\_3  &     -0.0558\sym{***}&     -0.0764\sym{***}&     -0.0999\sym{***}&     -0.0909\sym{***}&                     &                     &                     \\
            &    (0.0138)         &    (0.0147)         &    (0.0143)         &    (0.0147)         &                     &                     &                     \\
[1em]
cases\_4  &    -0.00274         &     -0.0255         &     -0.0394\sym{**} &     -0.0442\sym{**} &                     &                     &                     \\
            &    (0.0131)         &    (0.0132)         &    (0.0133)         &    (0.0142)         &                     &                     &                     \\
[1em]
deaths\_1 &                     &                     &                     &                     &      -0.559\sym{***}&      -0.570\sym{***}&      -0.590\sym{***}\\
            &                     &                     &                     &                     &    (0.0160)         &    (0.0160)         &    (0.0169)         \\
[1em]
deaths\_2 &                     &                     &                     &                     &      -0.248\sym{***}&      -0.256\sym{***}&      -0.284\sym{***}\\
            &                     &                     &                     &                     &    (0.0191)         &    (0.0188)         &    (0.0186)         \\
[1em]
deaths\_3 &                     &                     &                     &                     &      -0.102\sym{***}&     -0.0988\sym{***}&      -0.130\sym{***}\\
            &                     &                     &                     &                     &    (0.0160)         &    (0.0158)         &    (0.0156)         \\
[1em]
deaths\_4 &                     &                     &                     &                     &     -0.0326\sym{*}  &     -0.0243         &     -0.0337\sym{*}  \\
            &                     &                     &                     &                     &    (0.0141)         &    (0.0144)         &    (0.0139)         \\
[1em]
\_cons      &       1.472\sym{***}&       1.292\sym{***}&       1.479\sym{***}&       0.835\sym{***}&       1.225\sym{***}&       1.463\sym{***}&       1.595\sym{***}\\
            &    (0.0789)         &    (0.0819)         &    (0.0854)         &    (0.0925)         &     (0.105)         &     (0.105)         &     (0.115)         \\
\hline
$R^2$          &       0.282         &       0.269         &       0.278         &       0.259         &       0.294         &       0.297         &       0.313         \\
$R_{overall}^2$        &       0.187         &       0.183         &       0.164         &       0.205         &       0.227         &       0.232         &       0.227         \\
N           &        8815         &        8485         &        8142         &        7800         &        7377         &        7126         &        6861         \\
p           &           .         &           .         &           .         &           .         &           .         &           .         &           .         \\
\hline\hline
\multicolumn{8}{l}{\footnotesize Standard errors in parentheses}\\
\multicolumn{8}{l}{\footnotesize \sym{*} \(p<0.05\), \sym{**} \(p<0.01\), \sym{***} \(p<0.001\)}\\
\end{tabular}
}

\begin{tablenotes}
\small
\item[a]{Results for Fixed-Effects Estimation over 2020 and 2021 sample, aggregating cases (deaths) by residence area and using overall vaccination from OpenData SUS data set. $R^2$ denotes the R-squared, $R_{overall}^2$ the overall R-squared, $N$ the total number of observations used on the estimation and $p$ the F-test associated p-value.}
\end{tablenotes}
\end{threeparttable}
\end{adjustbox}
\end{table}

\begin{table}[H]
\centering
\caption{Estimation for Complete Sample (2020-2021) by Notification Area\label{tab:not2021}}
\begin{adjustbox}{max width= 13cm}
\begin{threeparttable}
{
\def\sym#1{\ifmmode^{#1}\else\(^{#1}\)\fi}
\begin{tabular}{l*{7}{c}}
\hline\hline
            &\multicolumn{1}{c}{(1)}&\multicolumn{1}{c}{(2)}&\multicolumn{1}{c}{(3)}&\multicolumn{1}{c}{(4)}&\multicolumn{1}{c}{(5)}&\multicolumn{1}{c}{(6)}&\multicolumn{1}{c}{(7)}\\
            &\multicolumn{1}{c}{cases}&\multicolumn{1}{c}{cases}&\multicolumn{1}{c}{cases}&\multicolumn{1}{c}{cases}&\multicolumn{1}{c}{deaths}&\multicolumn{1}{c}{deaths}&\multicolumn{1}{c}{deaths}\\
            &         m=1         &         m=2         &         m=3         &         m=4         &         m=2         &         m=3         &         m=4         \\
\hline
residential &     -0.0501\sym{***}&     -0.0455\sym{***}&     -0.0413\sym{***}&     -0.0274\sym{***}&     -0.0217\sym{***}&     -0.0495\sym{***}&     -0.0581\sym{***}\\
            &   (0.00417)         &   (0.00435)         &   (0.00443)         &   (0.00403)         &   (0.00474)         &   (0.00508)         &   (0.00578)         \\
[1em]
workplace   &     0.00103         &  -0.0000911         &   -0.000657         &  -0.0000294         &     0.00184         &     0.00195         &     0.00268         \\
            &   (0.00119)         &   (0.00122)         &   (0.00125)         &   (0.00127)         &   (0.00146)         &   (0.00157)         &   (0.00172)         \\
[1em]
1st\_dose    &     0.00457         &    -0.00839         &    0.000994         &    -0.00157         &    -0.00121         &    -0.00592         &     0.00102         \\
            &   (0.00465)         &   (0.00501)         &   (0.00512)         &   (0.00529)         &   (0.00651)         &   (0.00674)         &   (0.00692)         \\
[1em]
2nd\_dose    &    0.000531         &    -0.00689\sym{*}  &    -0.00227         &    -0.00172         &    -0.00182         &    -0.00714         &     0.00194         \\
            &   (0.00341)         &   (0.00310)         &   (0.00317)         &   (0.00355)         &   (0.00452)         &   (0.00436)         &   (0.00445)         \\
[1em]
n\_covid     &    0.000673\sym{***}&    0.000441\sym{**} &   0.0000867         &    0.000250         &    0.000450         &   0.0000454         &   -0.000158         \\
            &  (0.000174)         &  (0.000170)         &  (0.000186)         &  (0.000175)         &  (0.000268)         &  (0.000267)         &  (0.000243)         \\
[1em]
n\_prevention&    0.000307         &     0.00146         &     0.00128         &    0.000789         &    -0.00145         &     0.00148         &     0.00302\sym{**} \\
            &  (0.000777)         &  (0.000797)         &  (0.000804)         &  (0.000841)         &   (0.00113)         &   (0.00117)         &   (0.00115)         \\
[1em]
n\_fakenews  &     -0.0124\sym{*}  &     -0.0267\sym{***}&    -0.00355         &     0.00262         &     -0.0241\sym{***}&     -0.0165\sym{*}  &     -0.0240\sym{***}\\
            &   (0.00526)         &   (0.00519)         &   (0.00505)         &   (0.00518)         &   (0.00704)         &   (0.00724)         &   (0.00698)         \\
[1em]
n\_vaccines  &    0.000194         &   -0.000281         &    -0.00107\sym{***}&   -0.000483         &    0.000317         &   0.0000689         &   -0.000565         \\
            &  (0.000269)         &  (0.000288)         &  (0.000311)         &  (0.000332)         &  (0.000359)         &  (0.000400)         &  (0.000452)         \\
[1em]
gt\_covid    &   0.0000713         &   -0.000175\sym{***}&   -0.000330\sym{***}&   -0.000418\sym{***}&    0.000456\sym{***}&    0.000240\sym{***}&   0.0000844         \\
            & (0.0000438)         & (0.0000438)         & (0.0000445)         & (0.0000448)         & (0.0000654)         & (0.0000588)         & (0.0000613)         \\
[1em]
gt\_prevention&   0.0000499         &    0.000378\sym{***}&  -0.0000655         &  -0.0000361         &    0.000131         &   0.0000830         &  -0.0000845         \\
            & (0.0000942)         & (0.0000983)         & (0.0000936)         & (0.0000958)         &  (0.000129)         &  (0.000124)         &  (0.000130)         \\
[1em]
gt\_fakenews &   0.0000206         &   -0.000294\sym{*}  &  -0.0000171         &    0.000132         &   0.0000763         &    0.000125         &   -0.000273         \\
            &  (0.000133)         &  (0.000134)         &  (0.000132)         &  (0.000130)         &  (0.000179)         &  (0.000173)         &  (0.000178)         \\
[1em]
gt\_vaccines &   -0.000346\sym{***}&   -0.000150         &    0.000118         &    0.000337\sym{***}&   -0.000664\sym{***}&   -0.000463\sym{***}&   -0.000416\sym{***}\\
            & (0.0000887)         & (0.0000842)         & (0.0000876)         & (0.0000896)         &  (0.000115)         &  (0.000120)         &  (0.000124)         \\
[1em]
cases\_1  &      -0.486\sym{***}&      -0.494\sym{***}&      -0.500\sym{***}&      -0.502\sym{***}&                     &                     &                     \\
            &    (0.0114)         &    (0.0120)         &    (0.0123)         &    (0.0129)         &                     &                     &                     \\
[1em]
cases\_2  &      -0.230\sym{***}&      -0.253\sym{***}&      -0.263\sym{***}&      -0.260\sym{***}&                     &                     &                     \\
            &    (0.0117)         &    (0.0119)         &    (0.0125)         &    (0.0131)         &                     &                     &                     \\
[1em]
cases\_3  &      -0.116\sym{***}&      -0.130\sym{***}&      -0.148\sym{***}&      -0.153\sym{***}&                     &                     &                     \\
            &    (0.0109)         &    (0.0108)         &    (0.0112)         &    (0.0115)         &                     &                     &                     \\
[1em]
cases\_4  &     -0.0415\sym{***}&     -0.0572\sym{***}&     -0.0663\sym{***}&     -0.0716\sym{***}&                     &                     &                     \\
            &   (0.00931)         &   (0.00948)         &    (0.0100)         &    (0.0104)         &                     &                     &                     \\
[1em]
deaths\_1 &                     &                     &                     &                     &      -0.595\sym{***}&      -0.605\sym{***}&      -0.615\sym{***}\\
            &                     &                     &                     &                     &    (0.0118)         &    (0.0121)         &    (0.0121)         \\
[1em]
deaths\_2 &                     &                     &                     &                     &      -0.324\sym{***}&      -0.333\sym{***}&      -0.347\sym{***}\\
            &                     &                     &                     &                     &    (0.0139)         &    (0.0138)         &    (0.0138)         \\
[1em]
deaths\_3 &                     &                     &                     &                     &      -0.161\sym{***}&      -0.170\sym{***}&      -0.182\sym{***}\\
            &                     &                     &                     &                     &    (0.0128)         &    (0.0125)         &    (0.0125)         \\
[1em]
deaths\_4 &                     &                     &                     &                     &     -0.0802\sym{***}&     -0.0821\sym{***}&     -0.0851\sym{***}\\
            &                     &                     &                     &                     &   (0.00990)         &    (0.0101)         &    (0.0100)         \\
[1em]
\_cons      &       1.307\sym{***}&       1.121\sym{***}&       1.463\sym{***}&       0.963\sym{***}&       1.096\sym{***}&       1.256\sym{***}&       1.522\sym{***}\\
            &    (0.0614)         &    (0.0636)         &    (0.0691)         &    (0.0685)         &    (0.0842)         &    (0.0859)         &    (0.0863)         \\
\hline
$R^2$          &       0.265         &       0.267         &       0.266         &       0.260         &       0.299         &       0.309         &       0.312         \\
$R_{overall}^2$        &       0.189         &       0.200         &       0.170         &       0.201         &       0.244         &       0.243         &       0.224         \\
N           &       17168         &       16548         &       15933         &       15304         &       12312         &       11941         &       11553         \\
p           &           .         &           .         &           .         &           .         &           .         &           .         &           .         \\
\hline\hline
\multicolumn{8}{l}{\footnotesize Standard errors in parentheses}\\
\multicolumn{8}{l}{\footnotesize \sym{*} \(p<0.05\), \sym{**} \(p<0.01\), \sym{***} \(p<0.001\)}\\
\end{tabular}
}

\begin{tablenotes}
\small
\item[a]{Results for Fixed-Effects Estimation over 2020 and 2021 sample, aggregating cases (deaths) by notification area and using overall vaccination from OpenData SUS data set. $R^2$ denotes the R-squared, $R_{overall}^2$ the overall R-squared, $N$ the total number of observations used on the estimation and $p$ the F-test associated p-value.}
\end{tablenotes}
\end{threeparttable}
\end{adjustbox}
\end{table}

\begin{table}[H]
\centering
\caption{Estimation for Complete Sample (2020-2021) using SRAG Vaccination Data\label{tab:vac}}
\begin{adjustbox}{max width= 13cm}
\begin{threeparttable}
{
\def\sym#1{\ifmmode^{#1}\else\(^{#1}\)\fi}
\begin{tabular}{l*{7}{c}}
\hline\hline
            &\multicolumn{1}{c}{(1)}&\multicolumn{1}{c}{(2)}&\multicolumn{1}{c}{(3)}&\multicolumn{1}{c}{(4)}&\multicolumn{1}{c}{(5)}&\multicolumn{1}{c}{(6)}&\multicolumn{1}{c}{(7)}\\
            &\multicolumn{1}{c}{cases}&\multicolumn{1}{c}{cases}&\multicolumn{1}{c}{cases}&\multicolumn{1}{c}{cases}&\multicolumn{1}{c}{deaths}&\multicolumn{1}{c}{deaths}&\multicolumn{1}{c}{deaths}\\
            &         m=1         &         m=2         &         m=3         &         m=4         &         m=2         &         m=3         &         m=4         \\
\hline
residential &     -0.0552\sym{***}&     -0.0485\sym{***}&     -0.0387\sym{***}&     -0.0331\sym{***}&     -0.0321\sym{***}&     -0.0535\sym{***}&     -0.0599\sym{***}\\
            &   (0.00473)         &   (0.00504)         &   (0.00524)         &   (0.00510)         &   (0.00568)         &   (0.00593)         &   (0.00817)         \\
[1em]
workplace   &    0.000145         &  -0.0000122         &    0.000458         &   -0.000311         &    0.000358         &     0.00366         &     0.00248         \\
            &   (0.00135)         &   (0.00144)         &   (0.00155)         &   (0.00164)         &   (0.00177)         &   (0.00203)         &   (0.00258)         \\
[1em]
srag\_vac    &     -0.0614\sym{***}&     -0.0688\sym{***}&     -0.0773\sym{***}&     -0.0702\sym{***}&      0.0816\sym{***}&      0.0372\sym{***}&     -0.0187         \\
            &   (0.00727)         &   (0.00744)         &   (0.00824)         &   (0.00822)         &    (0.0104)         &    (0.0104)         &    (0.0106)         \\
[1em]
n\_covid     &    0.000277         &    0.000156         &    0.000278         &    0.000580\sym{**} &    0.000152         &    0.000177         &    0.000115         \\
            &  (0.000205)         &  (0.000204)         &  (0.000216)         &  (0.000193)         &  (0.000306)         &  (0.000306)         &  (0.000286)         \\
[1em]
n\_prevention&    0.000556         &    0.000229         &   -0.000201         &   -0.000811         &    -0.00107         &     0.00145         &     0.00193         \\
            &  (0.000846)         &  (0.000893)         &  (0.000940)         &  (0.000923)         &   (0.00137)         &   (0.00129)         &   (0.00133)         \\
[1em]
n\_fakenews  &     -0.0199\sym{***}&     -0.0182\sym{**} &   -0.000406         &    -0.00155         &     -0.0176\sym{*}  &     -0.0186\sym{*}  &     -0.0259\sym{**} \\
            &   (0.00540)         &   (0.00564)         &   (0.00552)         &   (0.00538)         &   (0.00764)         &   (0.00773)         &   (0.00806)         \\
[1em]
n\_vaccines  &   -0.000195         &   -0.000136         &   -0.000360         &   -0.000639         &    0.000188         &   -0.000557         &   -0.000164         \\
            &  (0.000308)         &  (0.000323)         &  (0.000349)         &  (0.000389)         &  (0.000410)         &  (0.000431)         &  (0.000521)         \\
[1em]
gt\_covid    &   0.0000932         &   -0.000120\sym{*}  &   -0.000308\sym{***}&   -0.000485\sym{***}&    0.000351\sym{***}&    0.000169\sym{*}  &    0.000151\sym{*}  \\
            & (0.0000513)         & (0.0000547)         & (0.0000560)         & (0.0000568)         & (0.0000789)         & (0.0000773)         & (0.0000752)         \\
[1em]
gt\_prevention&   0.0000561         &    0.000289\sym{*}  &   0.0000517         &    0.000138         &    0.000316         &    0.000130         &   -0.000173         \\
            &  (0.000118)         &  (0.000125)         &  (0.000132)         &  (0.000125)         &  (0.000176)         &  (0.000169)         &  (0.000162)         \\
[1em]
gt\_fakenews &   0.0000914         &  -0.0000894         &  0.00000143         &    0.000317\sym{*}  &    0.000228         &    0.000268         &   -0.000300         \\
            &  (0.000147)         &  (0.000152)         &  (0.000166)         &  (0.000158)         &  (0.000217)         &  (0.000216)         &  (0.000220)         \\
[1em]
gt\_vaccines &   -0.000271\sym{**} &   -0.000163         &   0.0000486         &    0.000334\sym{**} &   -0.000614\sym{***}&   -0.000267         &   -0.000594\sym{***}\\
            &  (0.000102)         &  (0.000104)         &  (0.000106)         &  (0.000111)         &  (0.000133)         &  (0.000144)         &  (0.000152)         \\
[1em]
cases\_1  &      -0.409\sym{***}&      -0.519\sym{***}&      -0.514\sym{***}&      -0.533\sym{***}&                     &                     &                     \\
            &    (0.0121)         &    (0.0127)         &    (0.0122)         &    (0.0130)         &                     &                     &                     \\
[1em]
cases\_2  &      -0.195\sym{***}&      -0.203\sym{***}&      -0.301\sym{***}&      -0.319\sym{***}&                     &                     &                     \\
            &    (0.0137)         &    (0.0141)         &    (0.0137)         &    (0.0148)         &                     &                     &                     \\
[1em]
cases\_3  &     -0.0900\sym{***}&     -0.0887\sym{***}&      -0.101\sym{***}&      -0.202\sym{***}&                     &                     &                     \\
            &    (0.0132)         &    (0.0139)         &    (0.0156)         &    (0.0156)         &                     &                     &                     \\
[1em]
cases\_4  &     -0.0331\sym{**} &     -0.0383\sym{***}&     -0.0307\sym{*}  &     -0.0375\sym{**} &                     &                     &                     \\
            &    (0.0109)         &    (0.0109)         &    (0.0120)         &    (0.0131)         &                     &                     &                     \\
[1em]
deaths\_1 &                     &                     &                     &                     &      -0.625\sym{***}&      -0.629\sym{***}&      -0.653\sym{***}\\
            &                     &                     &                     &                     &    (0.0132)         &    (0.0132)         &    (0.0137)         \\
[1em]
deaths\_2 &                     &                     &                     &                     &      -0.344\sym{***}&      -0.340\sym{***}&      -0.380\sym{***}\\
            &                     &                     &                     &                     &    (0.0149)         &    (0.0147)         &    (0.0153)         \\
[1em]
deaths\_3 &                     &                     &                     &                     &      -0.164\sym{***}&      -0.154\sym{***}&      -0.191\sym{***}\\
            &                     &                     &                     &                     &    (0.0141)         &    (0.0137)         &    (0.0143)         \\
[1em]
deaths\_4 &                     &                     &                     &                     &     -0.0591\sym{***}&     -0.0472\sym{***}&     -0.0776\sym{***}\\
            &                     &                     &                     &                     &    (0.0115)         &    (0.0120)         &    (0.0125)         \\
[1em]
\_cons      &       0.258\sym{**} &       0.371\sym{***}&       0.385\sym{***}&       0.583\sym{***}&       0.130         &       0.348\sym{***}&       0.441\sym{***}\\
            &    (0.0862)         &    (0.0699)         &    (0.0852)         &    (0.0771)         &    (0.0915)         &    (0.0919)         &     (0.107)         \\
\hline
$R^2$          &       0.220         &       0.259         &       0.246         &       0.265         &       0.314         &       0.312         &       0.324         \\
$R_{overall}^2$        &       0.119         &       0.161         &       0.148         &       0.181         &       0.272         &       0.258         &       0.244         \\
N           &       12761         &       12119         &       11435         &       10747         &        9577         &        9151         &        8666         \\
p           &   9.35e-258         &   2.05e-262         &   2.39e-254         &   4.63e-250         &   9.69e-257         &   7.69e-247         &   1.89e-242         \\
\hline\hline
\multicolumn{8}{l}{\footnotesize Standard errors in parentheses}\\
\multicolumn{8}{l}{\footnotesize \sym{*} \(p<0.05\), \sym{**} \(p<0.01\), \sym{***} \(p<0.001\)}\\
\end{tabular}
}

\begin{tablenotes}
\small
\item[a]{Results for Fixed-Effects Estimation over 2020 and 2021 sample, aggregating cases (deaths) by notification area and using vaccination data from the SRAG data set. In this case, the vaccination data reveals only series for individuals that has been vaccinated but entered on the hospital due to Covid-19 infection. $R^2$ denotes the R-squared, $R_{overall}^2$ the overall R-squared, $N$ the total number of observations used on the estimation and $p$ the F-test associated p-value.}
\end{tablenotes}
\end{threeparttable}
\end{adjustbox}
\end{table}

\begin{table}[H]
\centering
\caption{Estimation for Complete Sample (2020-2021) Using Dynamic Panel\label{tab:dynpan}}
\begin{adjustbox}{max width= 13cm}
\begin{threeparttable}
{
\def\sym#1{\ifmmode^{#1}\else\(^{#1}\)\fi}
\begin{tabular}{l*{7}{c}}
\hline\hline
            &\multicolumn{1}{c}{(1)}&\multicolumn{1}{c}{(2)}&\multicolumn{1}{c}{(3)}&\multicolumn{1}{c}{(4)}&\multicolumn{1}{c}{(5)}&\multicolumn{1}{c}{(6)}&\multicolumn{1}{c}{(7)}\\
            &\multicolumn{1}{c}{cases}&\multicolumn{1}{c}{cases}&\multicolumn{1}{c}{cases}&\multicolumn{1}{c}{cases}&\multicolumn{1}{c}{deaths}&\multicolumn{1}{c}{deaths}&\multicolumn{1}{c}{deaths}\\
            &         m=1         &         m=2         &         m=3         &         m=4         &         m=2         &         m=3         &         m=4         \\
\hline
cases\_1  &      -0.412\sym{***}&      -0.428\sym{***}&      -0.440\sym{***}&      -0.454\sym{***}&                     &                     &                     \\
            &    (0.0100)         &    (0.0100)         &    (0.0106)         &    (0.0110)         &                     &                     &                     \\
[1em]
cases\_2 &      -0.143\sym{***}&      -0.158\sym{***}&      -0.182\sym{***}&      -0.199\sym{***}&                     &                     &                     \\
            &    (0.0106)         &    (0.0111)         &    (0.0117)         &    (0.0118)         &                     &                     &                     \\
[1em]
cases\_3 &     -0.0295\sym{**} &     -0.0456\sym{***}&     -0.0671\sym{***}&     -0.0799\sym{***}&                     &                     &                     \\
            &    (0.0101)         &    (0.0106)         &    (0.0113)         &    (0.0114)         &                     &                     &                     \\
[1em]
cases\_4 &      0.0114         &     0.00399         &     -0.0169         &     -0.0177         &                     &                     &                     \\
            &   (0.00860)         &   (0.00907)         &   (0.00952)         &   (0.00947)         &                     &                     &                     \\
[1em]
residential &     -0.0466\sym{***}&     -0.0237\sym{***}&     -0.0169\sym{**} &     0.00121         &     -0.0438\sym{***}&     -0.0606\sym{***}&     -0.0659\sym{***}\\
            &   (0.00454)         &   (0.00473)         &   (0.00534)         &   (0.00478)         &   (0.00728)         &   (0.00759)         &   (0.00969)         \\
[1em]
workplace   &  -0.0000937         &     0.00102         &   -0.000859         &   -0.000981         &   -0.000798         &    -0.00361         &    -0.00481         \\
            &   (0.00121)         &   (0.00122)         &   (0.00145)         &   (0.00137)         &   (0.00201)         &   (0.00212)         &   (0.00256)         \\
[1em]
1st\_dose    &     0.00275         &     -0.0162\sym{**} &     -0.0101         &     0.00952         &     -0.0155         &     -0.0199\sym{**} &     -0.0145         \\
            &   (0.00507)         &   (0.00526)         &   (0.00516)         &   (0.00504)         &   (0.00792)         &   (0.00765)         &   (0.00744)         \\
[1em]
2nd\_dose    &      0.0187\sym{***}&      0.0137\sym{***}&      0.0220\sym{***}&     0.00884\sym{**} &    -0.00239         &     0.00650         &      0.0138\sym{**} \\
            &   (0.00312)         &   (0.00294)         &   (0.00301)         &   (0.00308)         &   (0.00411)         &   (0.00449)         &   (0.00455)         \\
[1em]
n\_covid     &    -0.00140\sym{***}&    -0.00108\sym{***}&   -0.000810\sym{***}&     0.00104\sym{***}&    -0.00119\sym{***}&    -0.00169\sym{***}&    -0.00166\sym{***}\\
            &  (0.000226)         &  (0.000232)         &  (0.000235)         &  (0.000230)         &  (0.000352)         &  (0.000328)         &  (0.000337)         \\
[1em]
n\_prevention&    0.000408         &     0.00116         &     0.00187\sym{*}  &    -0.00222\sym{**} &    -0.00130         &    0.000397         &     0.00261\sym{*}  \\
            &  (0.000752)         &  (0.000798)         &  (0.000767)         &  (0.000748)         &   (0.00117)         &   (0.00117)         &   (0.00111)         \\
[1em]
n\_fakenews  &     -0.0116\sym{*}  &  0.00000821         &    -0.00489         &     0.00311         &    -0.00951         &    -0.00893         &     -0.0146\sym{*}  \\
            &   (0.00459)         &   (0.00468)         &   (0.00459)         &   (0.00463)         &   (0.00667)         &   (0.00686)         &   (0.00681)         \\
[1em]
n\_vaccines  & -0.00000465         &   0.0000891         &   -0.000154         &    -0.00139\sym{***}&     0.00105\sym{*}  &     0.00113\sym{*}  &    0.000430         \\
            &  (0.000336)         &  (0.000349)         &  (0.000334)         &  (0.000348)         &  (0.000499)         &  (0.000490)         &  (0.000499)         \\
[1em]
gt\_covid    &    0.000180\sym{***}&   -0.000167\sym{***}&   -0.000326\sym{***}&   -0.000544\sym{***}&    0.000485\sym{***}&    0.000219\sym{**} &    0.000163\sym{*}  \\
            & (0.0000495)         & (0.0000463)         & (0.0000478)         & (0.0000492)         & (0.0000686)         & (0.0000675)         & (0.0000693)         \\
[1em]
gt\_prevention&   0.0000901         &    0.000171         &   -0.000175         &   -0.000179         &    0.000637\sym{***}&    0.000436\sym{**} &   0.0000299         \\
            &  (0.000102)         & (0.0000990)         & (0.0000993)         & (0.0000998)         &  (0.000150)         &  (0.000143)         &  (0.000147)         \\
[1em]
gt\_fakenews &   -0.000639\sym{***}&   -0.000418\sym{***}&  -0.0000529         &   0.0000535         &    0.000196         &   0.0000397         &   -0.000584\sym{***}\\
            &  (0.000121)         &  (0.000114)         &  (0.000125)         &  (0.000125)         &  (0.000168)         &  (0.000172)         &  (0.000168)         \\
[1em]
gt\_vaccines &   -0.000122         &   0.0000694         &    0.000269\sym{**} &    0.000728\sym{***}&   -0.000798\sym{***}&   -0.000454\sym{***}&   -0.000710\sym{***}\\
            & (0.0000844)         & (0.0000811)         & (0.0000834)         & (0.0000886)         &  (0.000115)         &  (0.000121)         &  (0.000127)         \\
[1em]
trend       &     -0.0256\sym{***}&     -0.0258\sym{***}&     -0.0283\sym{***}&     -0.0349\sym{***}&     -0.0104\sym{***}&     -0.0162\sym{***}&     -0.0142\sym{***}\\
            &   (0.00180)         &   (0.00209)         &   (0.00219)         &   (0.00232)         &   (0.00276)         &   (0.00318)         &   (0.00327)         \\
[1em]
deaths\_1 &                     &                     &                     &                     &      -0.538\sym{***}&      -0.534\sym{***}&      -0.552\sym{***}\\
            &                     &                     &                     &                     &    (0.0120)         &    (0.0122)         &    (0.0126)         \\
[1em]
deaths\_2&                     &                     &                     &                     &      -0.252\sym{***}&      -0.243\sym{***}&      -0.269\sym{***}\\
            &                     &                     &                     &                     &    (0.0141)         &    (0.0142)         &    (0.0146)         \\
[1em]
deaths\_3&                     &                     &                     &                     &     -0.0800\sym{***}&     -0.0693\sym{***}&     -0.0923\sym{***}\\
            &                     &                     &                     &                     &    (0.0132)         &    (0.0130)         &    (0.0131)         \\
[1em]
deaths\_4&                     &                     &                     &                     &     -0.0128         &    -0.00489         &     -0.0125         \\
            &                     &                     &                     &                     &    (0.0107)         &    (0.0108)         &    (0.0107)         \\
[1em]
\_cons      &       1.575\sym{***}&       1.865\sym{***}&       1.991\sym{***}&       2.078\sym{***}&       0.644\sym{***}&       1.373\sym{***}&       1.604\sym{***}\\
            &    (0.0811)         &    (0.0901)         &    (0.0971)         &     (0.107)         &     (0.129)         &     (0.135)         &     (0.143)         \\
\hline
N           &       19597         &       18830         &       18052         &       17266         &       12752         &       12368         &       11962         \\
p           &           0         &           0         &           0         &           0         &           0         &           0         &           0         \\
\hline\hline
\multicolumn{8}{l}{\footnotesize Standard errors in parentheses}\\
\multicolumn{8}{l}{\footnotesize \sym{*} \(p<0.05\), \sym{**} \(p<0.01\), \sym{***} \(p<0.001\)}\\
\end{tabular}
}

\begin{tablenotes}
\small
\item[a]{Results for Dynamic-Panel Estimation (Arellano-Bond using four lags) over 2020 and 2021 sample, aggregating cases (deaths) by residence area and using overall vaccination from OpenData SUS data set. $R^2$ denotes the R-squared, $R_{overall}^2$ the overall R-squared, $N$ the total number of observations used on the estimation and $p$ the F-test associated p-value.}
\end{tablenotes}
\end{threeparttable}
\end{adjustbox}
\end{table}

\begin{table}[H]
\centering
\caption{Estimation for Sub-Sample (2020) Using Dynamic Panel\label{tab:dynpan2}}
\begin{adjustbox}{max width= 13cm}
\begin{threeparttable}
{
\def\sym#1{\ifmmode^{#1}\else\(^{#1}\)\fi}
\begin{tabular}{l*{7}{c}}
\hline\hline
            &\multicolumn{1}{c}{(1)}&\multicolumn{1}{c}{(2)}&\multicolumn{1}{c}{(3)}&\multicolumn{1}{c}{(4)}&\multicolumn{1}{c}{(5)}&\multicolumn{1}{c}{(6)}&\multicolumn{1}{c}{(7)}\\
            &\multicolumn{1}{c}{cases}&\multicolumn{1}{c}{cases}&\multicolumn{1}{c}{cases}&\multicolumn{1}{c}{cases}&\multicolumn{1}{c}{deaths}&\multicolumn{1}{c}{deaths}&\multicolumn{1}{c}{deaths}\\
            &         m=1         &         m=2         &         m=3         &         m=4         &         m=2         &         m=3         &         m=4         \\
\hline
cases\_1  &      -0.409\sym{***}&      -0.401\sym{***}&      -0.399\sym{***}&      -0.415\sym{***}&                     &                     &                     \\
            &    (0.0118)         &    (0.0116)         &    (0.0119)         &    (0.0118)         &                     &                     &                     \\
[1em]
cases\_2 &      -0.115\sym{***}&      -0.105\sym{***}&      -0.110\sym{***}&      -0.120\sym{***}&                     &                     &                     \\
            &    (0.0141)         &    (0.0140)         &    (0.0139)         &    (0.0139)         &                     &                     &                     \\
[1em]
cases\_3 &    -0.00379         &     0.00492         &    -0.00397         &     -0.0115         &                     &                     &                     \\
            &    (0.0125)         &    (0.0129)         &    (0.0125)         &    (0.0123)         &                     &                     &                     \\
[1em]
cases\_4 &      0.0104         &      0.0206         &      0.0144         &     0.00909         &                     &                     &                     \\
            &    (0.0104)         &    (0.0106)         &    (0.0105)         &    (0.0102)         &                     &                     &                     \\
[1em]
residential &     -0.0158\sym{*}  &    0.000758         &     -0.0182\sym{**} &     -0.0385\sym{***}&     0.00426         &     -0.0329\sym{***}&     -0.0280\sym{**} \\
            &   (0.00641)         &   (0.00624)         &   (0.00593)         &   (0.00521)         &   (0.00969)         &   (0.00869)         &   (0.00974)         \\
[1em]
workplace   &   -0.000629         &     0.00230         &    0.000361         &    -0.00562\sym{***}&     0.00557\sym{*}  &    -0.00106         &    -0.00491         \\
            &   (0.00167)         &   (0.00167)         &   (0.00162)         &   (0.00162)         &   (0.00260)         &   (0.00245)         &   (0.00289)         \\
[1em]
n\_covid     &   -0.000607         &   0.0000239         &    -0.00106\sym{*}  &    0.000839\sym{*}  &   -0.000499         &   -0.000407         &   -0.000843         \\
            &  (0.000413)         &  (0.000409)         &  (0.000436)         &  (0.000346)         &  (0.000565)         &  (0.000570)         &  (0.000538)         \\
[1em]
n\_prevention&      0.0104\sym{***}&     0.00484\sym{**} &     0.00330\sym{*}  &     0.00448\sym{**} &      0.0111\sym{***}&     0.00831\sym{***}&     0.00543\sym{**} \\
            &   (0.00169)         &   (0.00163)         &   (0.00153)         &   (0.00136)         &   (0.00236)         &   (0.00215)         &   (0.00196)         \\
[1em]
n\_fakenews  &     0.00426         &    -0.00733         &     0.00890         &    -0.00600         &    -0.00349         &     0.00313         &    -0.00406         \\
            &   (0.00720)         &   (0.00720)         &   (0.00764)         &   (0.00750)         &    (0.0111)         &    (0.0111)         &    (0.0107)         \\
[1em]
n\_vaccines  &    -0.00755\sym{***}&     0.00481\sym{*}  &     0.00101         &     0.00877\sym{***}&     -0.0156\sym{***}&    -0.00515         &    -0.00769\sym{*}  \\
            &   (0.00205)         &   (0.00209)         &   (0.00241)         &   (0.00224)         &   (0.00308)         &   (0.00351)         &   (0.00307)         \\
[1em]
gt\_covid    &  -0.0000222         &   -0.000110         &   0.0000131         &   -0.000400\sym{***}&    0.000272\sym{*}  &    0.000128         &    0.000333\sym{**} \\
            & (0.0000788)         & (0.0000768)         & (0.0000811)         & (0.0000746)         &  (0.000112)         &  (0.000112)         &  (0.000124)         \\
[1em]
gt\_prevention&    0.000294         &    0.000522\sym{***}&   -0.000116         &    0.000775\sym{***}&    0.000472         &    0.000604\sym{*}  &  -0.0000311         \\
            &  (0.000164)         &  (0.000157)         &  (0.000158)         &  (0.000152)         &  (0.000243)         &  (0.000238)         &  (0.000248)         \\
[1em]
gt\_fakenews &   -0.000105         &   -0.000434\sym{**} &   -0.000276         & -0.00000728         &    0.000233         &    0.000204         &   -0.000930\sym{***}\\
            &  (0.000178)         &  (0.000164)         &  (0.000161)         &  (0.000148)         &  (0.000237)         &  (0.000223)         &  (0.000221)         \\
[1em]
gt\_vaccines &    0.000154         &   -0.000359         &    0.000428\sym{*}  &    0.000568\sym{*}  &   -0.000210         &   -0.000587\sym{*}  &   -0.000167         \\
            &  (0.000209)         &  (0.000200)         &  (0.000214)         &  (0.000225)         &  (0.000312)         &  (0.000292)         &  (0.000337)         \\
[1em]
trend       &    -0.00798\sym{*}  &    -0.00977\sym{**} &     -0.0204\sym{***}&    -0.00961\sym{**} &     0.00424         &    -0.00479         &    -0.00921         \\
            &   (0.00341)         &   (0.00326)         &   (0.00337)         &   (0.00328)         &   (0.00485)         &   (0.00490)         &   (0.00491)         \\
[1em]
deaths\_1 &                     &                     &                     &                     &      -0.548\sym{***}&      -0.540\sym{***}&      -0.530\sym{***}\\
            &                     &                     &                     &                     &    (0.0167)         &    (0.0164)         &    (0.0165)         \\
[1em]
deaths\_2&                     &                     &                     &                     &      -0.256\sym{***}&      -0.244\sym{***}&      -0.229\sym{***}\\
            &                     &                     &                     &                     &    (0.0198)         &    (0.0195)         &    (0.0200)         \\
[1em]
deaths\_3&                     &                     &                     &                     &     -0.0450\sym{*}  &     -0.0369\sym{*}  &     -0.0212         \\
            &                     &                     &                     &                     &    (0.0178)         &    (0.0180)         &    (0.0186)         \\
[1em]
deaths\_4&                     &                     &                     &                     &    -0.00108         &     0.00328         &      0.0124         \\
            &                     &                     &                     &                     &    (0.0138)         &    (0.0137)         &    (0.0144)         \\
[1em]
\_cons      &       0.255\sym{*}  &       0.205         &       0.657\sym{***}&       0.459\sym{***}&      -0.401\sym{*}  &       0.210         &       0.309         \\
            &     (0.124)         &     (0.123)         &     (0.126)         &     (0.121)         &     (0.185)         &     (0.179)         &     (0.191)         \\
\hline
N           &       15270         &       15220         &       15183         &       15153         &        7362         &        7361         &        7362         \\
p           &           0         &           0         &           0         &           0         &   1.11e-298         &   3.89e-290         &   3.71e-285         \\
\hline\hline
\multicolumn{8}{l}{\footnotesize Standard errors in parentheses}\\
\multicolumn{8}{l}{\footnotesize \sym{*} \(p<0.05\), \sym{**} \(p<0.01\), \sym{***} \(p<0.001\)}\\
\end{tabular}
}

\begin{tablenotes}
\small
\item[a]{Results for Dynamic-Panel Estimation (Arellano-Bond using four lags) over the restricted 2020 sample, aggregating cases (deaths) by residence area and using overall vaccination from OpenData SUS data set. $R^2$ denotes the R-squared, $R_{overall}^2$ the overall R-squared, $N$ the total number of observations used on the estimation and $p$ the F-test associated p-value.}
\end{tablenotes}
\end{threeparttable}
\end{adjustbox}
\end{table}

\end{document}